\documentclass[twocolumn]{aastex62}
\usepackage{bm}
\newcommand{\bjdtdb}{${\rm {BJD_{TDB}}}$}
\newcommand{\feh}{{\left[{\rm Fe}/{\rm H}\right]}}
\newcommand{\teff}{{T_{\rm eff}}}

\newcommand{\msun}{${\rm M}_\Sun$}
\newcommand{\rsun}{${\rm R}_\Sun$}

\newcommand{\mj}{${\,{\rm M}_{\rm J}}$}
\newcommand{\rj}{${\,{\rm R}_{\rm J}}$}

\newcommand{\three}{3.6\,$\mu$m\ }
\newcommand{\four}{4.5\,$\mu$m\ }
\newcommand{\threealt}{3.6\,$\mu$m}
\newcommand{\fouralt}{4.5\,$\mu$m}
\newcommand{\um}{$\mu$m}
\newcommand{\water}{$\mathrm{H}_2\mathrm{O}$\ }
\newcommand{\methane}{$\mathrm{CH}_4$\ }

\newcommand{\carbonmonoxide}{$\mathrm{CO}$\ }

\begin{document}

\title{A Significant Over-Luminosity in the Transiting Brown Dwarf CWW 89Ab}

\author{Thomas G.\ Beatty}
\affiliation{Department of Astronomy \& Astrophysics, The Pennsylvania State University, 525 Davey Lab, University Park, PA 16802; tbeatty@psu.edu}
\affiliation{Center for Exoplanets and Habitable Worlds, The Pennsylvania State University, 525 Davey Lab, University Park, PA 16802}
\author{Caroline V. Morley}
\affiliation{Department of Astronomy, Harvard University, 60 Garden Street, Cambridge, MA 02138, USA}
\author{Jason L. Curtis}
\altaffiliation{NSF Astronomy \& Astrophysics Postdoctoral Fellow}
\affiliation{Department of Astronomy, Columbia University, 550 West 120th Street, New York, NY 10027, USA}
\author{Adam Burrows}
\affiliation{Department of Astrophysical Sciences, Princeton University, Princeton, NJ 08544, USA}
\author{James R. A. Davenport}
\altaffiliation{NSF Astronomy \& Astrophysics Postdoctoral Fellow}
\altaffiliation{DIRAC Fellow}
\affiliation{Department of Physics \& Astronomy, Western Washington 
University, 516 High St., Bellingham, WA 98225, USA}
\affiliation{Department of Astronomy, University of Washington, Seattle, 
WA 98195, USA}
\author{Benjamin T. Montet}
\altaffiliation{NASA Sagan Postdoctoral Fellow}
\affiliation{Department of Astronomy and Astrophysics, University of Chicago, 5640 S. Ellis Ave., Chicago, IL 60637, USA}

\shorttitle{Over-Luminosity in CWW 89Ab}
\shortauthors{Beatty et al.}


\begin{abstract}
We observed eclipses of the transiting brown dwarf CWW 89Ab at \three and \four using Spitzer/IRAC. The CWW 89 binary system is a member of the $3.0\pm0.25$ Gyr-old open cluster Ruprecht 147, and is composed of a Sun-like primary and an early M-dwarf secondary separated by a projected distance of 25 AU. CWW 89Ab has a radius of $0.937\pm0.042$\,\rj\ and a mass of $36.5\pm0.1$\,\mj, and is on a 5.3 day orbit about CWW 89A with a non-zero eccentricity of $e=0.19$ \citep{curtis2016}. We strongly detect the eclipses of CWW 89Ab in both Spitzer channels as $\delta_{3.6}=1147\pm213$\,ppm and $\delta_{4.5}=1097\pm225$\,ppm after correcting for the dilution from CWW 89B. After accounting for the irradiation that CWW 89Ab receives from its host star, these measurements imply that the brown dwarf has an internal luminosity of $\log(\mathrm{L_{bol}/\mathrm{L}_\odot})=-4.19\pm0.14$. This is 16 times, or $9.3\,\sigma$, higher than model predictions given the known mass, radius, and age of CWW 89Ab. As we discuss, this over-luminosity is not explainable by an inaccurate age determination, additional stellar heating, nor tidal heating. Instead, we suggest that the anomalous luminosity of CWW 89Ab is caused by a dayside temperature inversion -- though a significant error in the evolutionary models is also a possibility. Importantly, a temperature inversion would require a super-stellar C/O ratio in CWW 89Ab's atmosphere. If this is indeed the case, it implies that CWW 89Ab is a 36.5\mj\ object that formed via core accretion processes. Finally, we use our measurement of CWW 89Ab’s orbital eccentricity, improved via these observations, to constrain the tidal quality factors of the brown dwarf and the host star CWW 89A to be $Q_{BD}>10^{4.15}$ and $Q_*>10^{9}$, respectively.
\end{abstract}

\section{Introduction}

Models for the radius and luminosity evolution of brown dwarfs are poorly constrained by observations. The basic problem is that while we have hundreds of precise luminosity measurements -- and ages for those objects in clusters or moving groups -- we have almost no independent mass, radius, age, and luminosity measurements for individual objects. Evolutionary models are thus driven to use use the entangled results of ensemble broadband color measurements to understand the underlying physical properties of mass and radius, and how they determine how brown dwarfs' luminosities change with age \cite[e.g.][]{sm08}. 

One method to attack the lack of independent masses, radii, and ages for brown dwarfs is to use transiting brown dwarfs. All of these objects have measured masses from radial velocities and radii measured using photometry, and currently there are 16 brown dwarfs known to transit main sequence stars (including CWW 89Ab). Since the ages of the host stars are generally difficult to determine, all but four of these objects have poorly constrained ages, and have proven difficult to use as evolutionary model inputs. Relating the transiting brown dwarfs to their field brethren is further complicated by the fact that most of them are on very short (a few days long) orbits about their host stars, and are therefore heavily irradiated on their daysides.

Nevertheless, atmospheric characterization of transiting brown dwarfs has already provided interesting information about brown dwarfs' physical properties. To date, only two transiting objects have had their dayside atmospheres characterized: KELT-1b \citep{beatty2014,beatty2017} and LHS 6343C \citep{montet2016}. Intriguingly, though KELT-1b is heated by stellar irradiation to temperatures thousands of degrees Kelvin above what would be its effective temperature as an isolated field object, the $H$-band eclipse spectrum of KELT-1b's dayside measured by \cite{beatty2017} exactly matches that of a 3250\,K M5 dwarf. By comparison, planetary-mass hot Jupiters at the same irradiation levels all show isothermal or ``inverted'' eclipse spectra. This difference indicates that despite the heavy external irradiation received by KELT-1b, its high surface gravity plays a dominant role in setting the dayside atmospheric structure \citep{beatty2017}. 

Despite this spectral similarity to a field brown dwarf, the radius of KELT-1b is 1.2 times larger than predicted by models. While this level of radius of inflation would be unremarkable in a hot Jupiter \citep{laughlin2011}, the higher mass of KELT-1b means that puffing up its atmosphere requires approximately 30 times more energy than for a hot Jupiter. Given the perennial difficulty in explaining the mechanism behind hot Jupiter radius inflation \citep{baraffe2014}, the radius inflation in KELT-1b is one indication that brown dwarf evolutionary models are not completely correct. \cite{stassun2012} have suggested that the radii of brown dwarfs may be inflated via magnetic effects, though we unfortunately have no chromospheric activity measurements for any of the transiting brown dwarfs.

The non-transiting brown dwarf WD0137-349B also suggests discrepancies in the models. The overall system is composed of a white dwarf orbited by a brown dwarf companion, and \cite{casewell2015} observed phase curves of the brown dwarf in the NIR and all four Spitzer/IRAC channels. \cite{casewell2015} found that the dayside of WD0137-349B was three times more luminous than predicted in $K$ and at \fouralt. They attributed this over-luminosity to $\mathrm{H}_2$ florescence or $\mathrm{H}_3^+$ emission on WD0137-349B's dayside. Though the authors believe one of these two effects is the cause of the over-luminosity, they note that there are large uncertainties in the florescence models, and that there may be some underlying issue with evolutionary models.

A final indicator of inaccuracies in models of luminosity evolution comes from brown dwarfs with astrometric mass (but no radius) measurements. Observations of the brown dwarf--brown dwarf binaries HD 130948BC \citep{dupuy2009} and Gl 417BC \citep{dupuy2014} show that all four suffer from what those authors dubbed the ``sub-stellar luminosity problem": given their well-measured masses and ages they are all approximately twice as luminous as predicted by evolutionary models. 

\begin{figure*}[t]
\vskip 0.00in
\includegraphics[width=1.0\linewidth,clip]{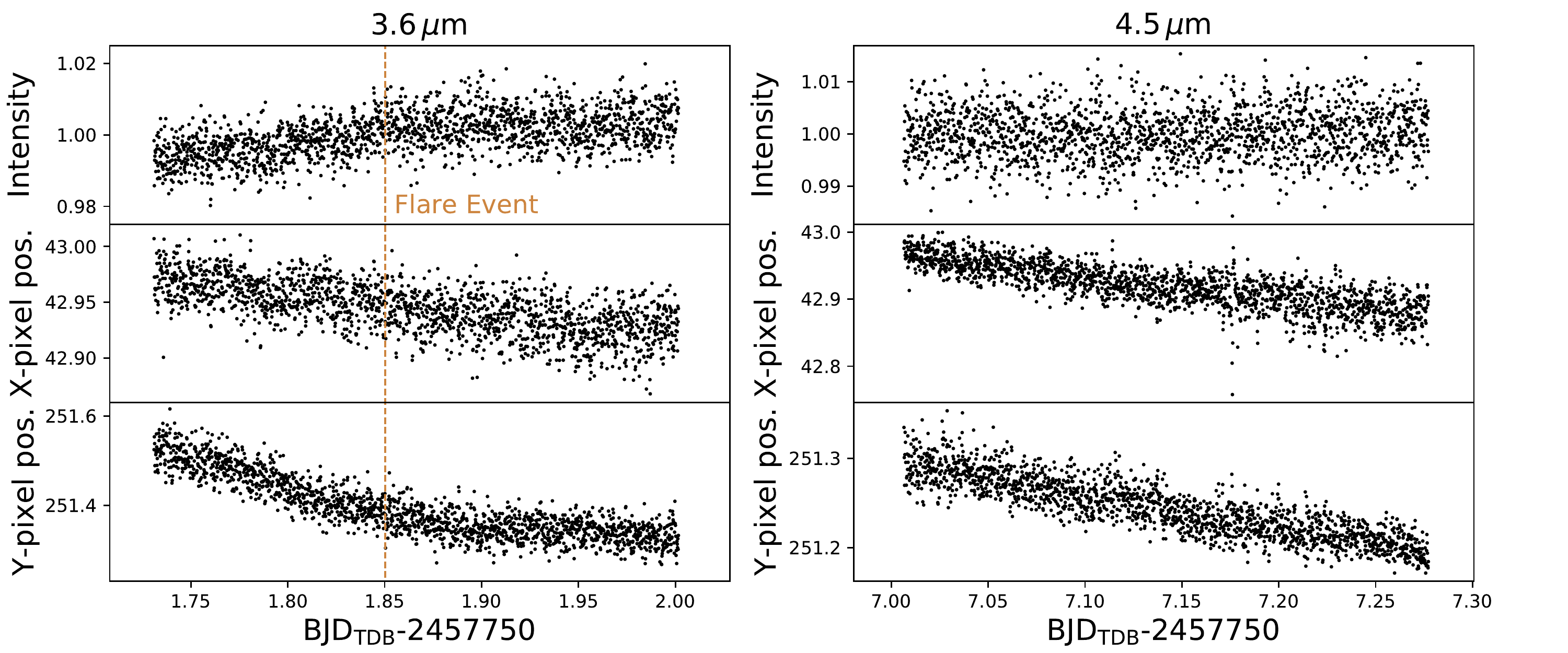}
\vskip -0.0in
\caption{The raw Spitzer photometry of CWW 89 in both bands (top panels), as well as the measured x-pixel and y-pixel positions of the star, all as a function of time. For the \three data, we have marked the time of the apparent stellar flare in gold. We discuss this event and our modeling of it in Section 3.3, but note that there is no corresponding jump in either of the position measurements that could explain the jump we see in the photometry.}
\label{rawplot}
\end{figure*}

As a notable counterpoint, however, the transiting brown dwarf LHS 6343C appears to agree with existing evolutionary models. \cite{montet2016} measured the Spitzer/IRAC eclipses (and hence the color) of the dayside of LHS 6343C and determined that the broadband color and the measured mass and radius for LHS 6343C were all consistent with a system age of 5.0 Gyr and the expected mid-T spectral type.

Note though, that for none of these objects have independently measured masses, radii, ages, and luminosities. That means the atmospheric ``hat-trick,'' as it were, remains to be made.

The recently discovered transiting brown dwarf CWW\footnote{The Curtis-Wolfgang-Wright (CWW) identification number for this star comes from the catalog of Ruprecht-147 stars in \cite{curtis2013}} 89Ab (EPIC 219388192b) provides just such an opportunity. CWW 89Ab was first identified by \cite{curtis2016}, and independently discovered by \cite{nowak2017}, who refined the system's physical parameters. The Kepler-K2 transit photometry and radial velocity observations show that CWW 89Ab has a radius of $0.937\pm0.042$\,\rj\ and a mass of $36.5\pm0.1$\,\mj. The CWW 89 binary stellar system is a member of the Ruprecht 147 open cluster, which provides us with a system age of $3.0\pm0.25$\, Gyr \citep{curtis2013}. CWW 89Ab is on a relatively long, 5.3 day, orbit about the Solar-twin CWW 89A, which means that its the zero-albedo, complete redistribution temperature is comparable to its expected effective temperature from internal heat alone (1150\,K vs. 850\,K). 

\begin{figure*}[th]
\vskip 0.00in
\includegraphics[width=1.0\linewidth,clip]{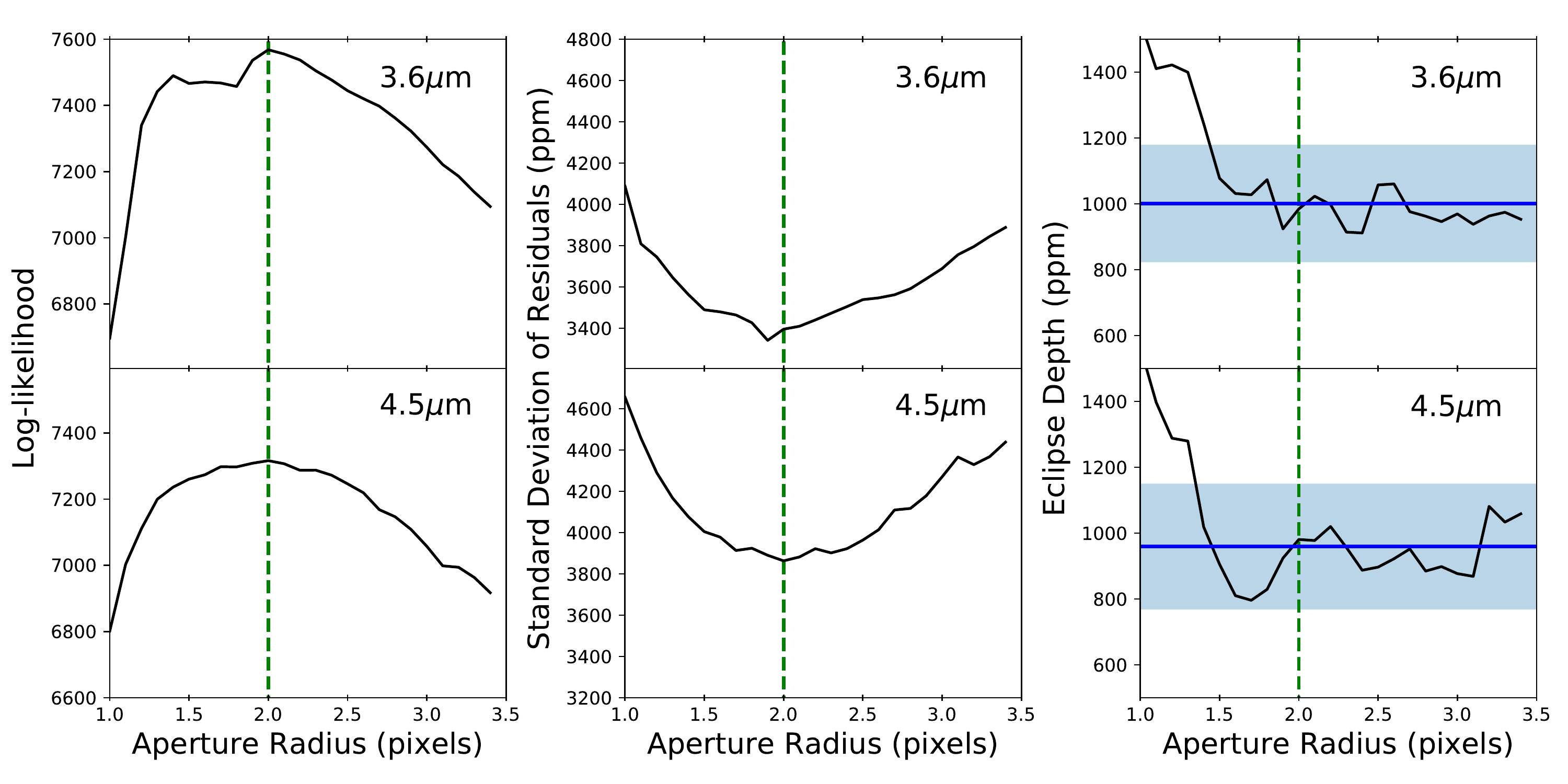}
\vskip -0.0in
\caption{The effect of differing extraction aperture radii on the fits to the Spitzer photometry. We selected an aperture of 2.0 pixels as the optimum, based on it having the highest log-likelihood and lowest residual RMS in our fits. As shown by the rightmost panels, the exact choice of aperture size does not significantly effect our results: the variation of the measured eclipse depths in both channels (black lines) as a function of aperture size does not exceed the $1\,\sigma$ uncertainties we find in our final fit (blue shaded regions) -- excepting for small apertures.}
\label{apervary}
\end{figure*}

CWW 89Ab is thus a brown dwarf with independent mass, radius, and age measurements -- and one that is not strongly irradiated by its host star. We wished to characterize its atmosphere, both to see how the atmosphere was effected by stellar irradiation, and to estimate the actual internal luminosity of the brown dwarf to compare to predictions from evolutionary models.

\section{Observations and Data Reduction}

We observed two secondary eclipses of CWW 89Ab using the \textsc{irac} camera on Spitzer Space Telescope. The first observation was at \three on UT 2016 December 29, and the second eclipse observation was at \four on UT 2017 January 03. For both eclipses we used an initial 30 minute observing sequence before the science observations to allow the spacecraft's pointing to settle and to allow the initial flux ramp-up commonly seen in these types of observations to reach a steady-state. We did not use these two initial sets of observations in our analysis. The science observing sequences began immediately following the 30 minute settling sequences, and both lasted for 6.93 hours.

Since CWW 89 is relatively faint ($K=10.666$), we used 12 second exposures at both \three and 4.5\um. This exposure length required taking full frame images with IRAC in both channels. We used Spitzer's PCRS Peak-up mode to place CWW 89 on the ``sweet-spot'' in the upper left corner of the detectors and to further stabilize the spacecraft's pointing. We used the nearby star HD 180514 as the peak-up target. The 12 second exposure time gave us 1860 full frame images in each band for use in our science analysis.

We began our image reduction process from the basic calibrated data (\textsc{bcd}) images provided to us by the Spitzer Science Center. We first calculated the mid-exposure time for each image. We did so by taking the \textsc{bmjd\_objs} header values for each exposure's start time, converted this to \bjdtdb, and then added one half the exposure time. We calculated the exposure time by using the \textsc{aintbeg} header values for the start time of each exposure, and the \textsc{atimeend} header values for the end time of each exposure and differencing the two.

We next took 40 pixel by 40 pixel subframes centered on CWW 89 out of the full frame \textsc{BCD} images to use in the rest of the reduction and lightcurve extraction process. To improve our measurements of the background level in each image and the location of CWW 89, we first eliminated bad pixels and cosmic ray hits. We did so by taking the timeseries for each individual pixel and performing three iterative rounds of $5\,\sigma$ clipping to identify images when that particular pixel suffered a potential cosmic ray hit. We replaced the pixel values in the outlier images with the median value of the pixel's clipped timeseries. We used these bad pixel corrected images only to measure the background level and CWW 89's position; the actual photometry in both bands was extracted from the uncorrected images.

To estimate the background level in each subframe, we masked out a central 20 pixel-square box around CWW 89 and took the median of the rest of the subframe. We measured the average background level as 0.08\,e$^-$\,pix$^{-1}$ at \three and 0.2\,e$^-$\,pix$^{-1}$ at 4.5\um. This was 0.02\% of CWW 89's flux at \three and 0.08\% of CWW 89's flux at 4.5\um, respectively. We measured the position of CWW 89 in each subframe by fitting a two-dimensional Gaussian to the bad pixel corrected, background-subtracted images. 

We then extracted photometry for CWW 89 in both bands using a circular aperture centered on the star's measured position in a subframe. We used the original \textsc{bcd} images with the background subtracted off. In each channel we extracted photometry for a range of fixed aperture radii from 1.0 to 3.5 pixels, in steps of 0.1 pixels. For reference, the average full-width half-maximum of CWW 89's point spread function was 1.61 pixels at 3.6\um, and 1.55 pixels at 4.5\um. We estimated the measurement uncertainty on the extracted fluxes by assuming pure photon-shot noise.

The first fifteen minutes of our initially extracted photometry showed a clear ramp-effect in both channels, which is a common feature of Spitzer/\textsc{irac} lightcurves. This occurred in the first fifteen minutes of the science observing sequence, after the initial thirty minute pre-flash. We therefore trimmed the first 75 images in each channel's lightcurve from our analysis. We also performed a single round of $5\,\sigma$ clipping on our extracted photometry. Together with the points trimmed to remove the ramp effect, this left us with 1776 flux measurements at 3.6\um, and 1768 measurements at 4.5\um\ (Figure \ref{rawplot}).  

\begin{deluxetable*}{lllc}
\tablecaption{Apparent Magnitudes of the CWW 89 Stellar System}
\tablehead{\colhead{~~~Parameter} & \colhead{Description} & \colhead{Value} & \colhead{Reference}}
\startdata
$B$\dotfill & APASS $B$ magnitude & $13.284\pm0.020$ & \cite{ucac4} \\
$V$\dotfill & APASS $V$ magnitude & $12.535\pm0.020$ & \cite{ucac4} \\
$r$\dotfill & APASS $r$ magnitude & $12.35\pm0.02$ & \cite{ucac4} \\
$i$\dotfill & APASS $i$ magnitude & $12.11\pm0.05$ & \cite{ucac4} \\
$J$\dotfill   & 2MASS $J$ magnitude   & $11.073\pm0.023$ & \cite{2mass} \\
$H$\dotfill   & 2MASS $H$ magnitude   & $10.734\pm0.021$ & \cite{2mass} \\
$K_S$\dotfill & 2MASS $K_S$ magnitude & $10.666\pm0.021$ & \cite{2mass} \\
\textit{W1}\dotfill & WISE magnitude\dotfill  & $10.583\pm0.023$ & \cite{wise} \\
\textit{W2}\dotfill & WISE magnitude\dotfill  & $10.611\pm0.020$ & \cite{wise} \\
\textit{Spitz. \threealt}\dotfill & Spitzer magnitude\dotfill  & $10.455\pm0.015$ & This work \\
\textit{Spitz. \fouralt}\dotfill & Spitzer magnitude\dotfill  & $10.485\pm0.015$ & This work 
\enddata
\label{tab:seddata}
\end{deluxetable*}

We chose the optimum photometric extraction aperture by running our fitting process, which is described in Section 3 on the lightcurves generated by each aperture. Though in our final results we fit the \three and \four data simultaneously, here we fit each channel independently. We judged the optimum photometric aperture to be the one which give the highest log-likelihood fit with the lowest scatter in residuals. As shown in Figure \ref{apervary}, this occurred in both channels at an aperture size of 2.0 pixels. Note too that the rightmost  panels of Figure \ref{apervary} show the secondary eclipse depths measured at each aperture for each channel, and that aside from very small aperture sizes the variation in depth as a function of aperture size is lower than our final depth uncertainties. This indicates that our precise choice of aperture size does not significantly affect our results. 

\subsection{Absolute Photometry of CWW 98}

Besides measuring the secondary eclipse of CWW 89Ab, we also used our \three and \four lightcurves of the CWW 89 system to measure its apparent and absolute magnitudes in the \textsc{irac} bands. Using the aperture corrections listed in Table 4.7 of the \textsc{irac} Instrument Handbook, the average measured fluxes from CWW 89 in our optimum photometric apertures was $49.185\pm0.010$\,e$^{-}$\,s$^{-1}$ at \three and $26.080\pm0.006$\,e$^{-}$\,s$^{-1}$ at 4.5\um. This corresponds to fluxes of $18,464.9\pm3.5$\,$\mu$Jy at \three and $11,490.6\pm2.4$\,$\mu$Jy at 4.5\um, and to apparent magnitudes of $[3.6]=10.455\pm0.015$ and $[4.5]=10.485\pm0.015$. We increased the uncertainties on these two magnitude measurement to account for the uncertainty in Spitzer's photometric zero-points \citep{carey2012}.

The Gaia DR2 parallax to the CWW 89 system is $3.251\pm0.049$\,mas \citep{gaiadr2}, which equates to a distance modulus of $7.44\pm0.08$. The absolute magnitudes of the CWW 89 system in the \textsc{irac} bandpasses are thus $M_{[3.6]}=3.02\pm0.08$ and $M_{[4.5]}=3.05\pm0.08$. Note, though, that this is for the CWW 89 system, not CWW 89A itself. After applying the dilution correction we determined in the next subsection, we find that CWW 89A has apparent magnitudes of $[3.6]=10.626\pm0.018$ and $[4.5]=10.650\pm0.019$, and absolute magnitudes of $M_{[3.6]}=3.19\pm0.08$ and $M_{[4.5]}=3.21\pm0.08$

\subsection{The CWW 89 Binary System and Eclipse Dilution Correction}

Adaptive optics (AO) images of CWW 89 taken using Keck/NIRC2 identified a probable binary companion to the primary star in the CWW 89 system \citep{curtis2018}. The companion star, which we will refer to as CWW 89B, is 81 mas from CWW 89A and is dimmer, at $\Delta K=2.238\pm0.025$. As described in \cite{curtis2018}, the brightness difference between the two stars implies that CWW 89B is an early M-dwarf with a mass of about 0.5\msun, and is at a projected distance of 24.9 AU from CWW 89A if the two stars are associated. \cite{curtis2018} concluded that the two stars are most likely gravitationally bound, since the presence of CWW 89B as a binary companion would explain a 4 km s$^{-1}$ discrepancy between the measured systemic velocity of CWW 89A and the average systemic velocity of the other R147 members.

The two stars were not resolved in any of our Spitzer images. We therefore needed to determine a dilution correction to account for the presence of CWW 89B in the eclipse photometry. Note that we are certain that transiting object in the CWW 89 must be in orbit about CWW 89A -- and thus is a brown dwarf -- since if the transits were occurring around CWW 89B the companion would have to be a late M-dwarf and would show much deeper secondary eclipses than we observe. Additionally, the radial velocity observations in \cite{curtis2018} clearly show the Doplper motion of the spectral features from CWW 89A.

\begin{figure}[t]
\vskip 0.00in
\includegraphics[width=1.1\linewidth,clip]{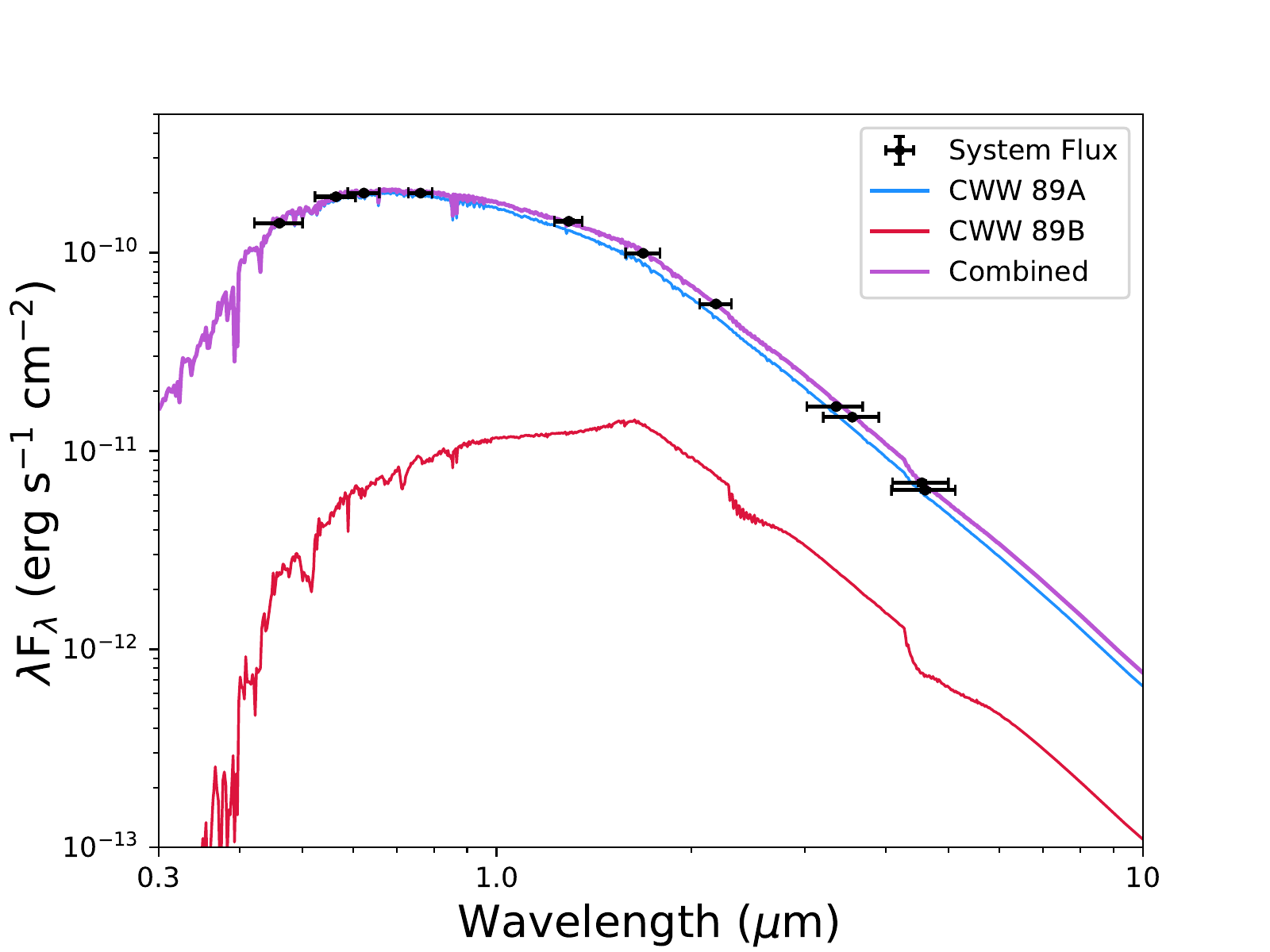}
\vskip -0.0in
\caption{The two stars CWW 89A and 89B are not resolved in our Spitzer images. By fitting the catalog magnitudes for the CWW 89 system (green points) and knowing the spectroscopic properties of CWW 89A (light orange line), we are able to estimate the SED of CWW 89B (dark orange line). This in turn allows us to estimate a dilution correction to account for the presence of CWW 89B, as well as estimate CWW 89B's properties (Table 2).}
\label{sed}
\vskip -0.1in
\end{figure}

To calculate a dilution correction for CWW 89B in the Spitzer bandpasses, we fit the spectral energy distribution (SED) of the combined CWW 89 system. Since we know the spectroscopic effective temperature and surface gravity of CWW 89A, this combined fit allowed us to estimate the individual SED of CWW 89B. We fit the combined SED of the CWW 89 system to the observed $B, V$, $r$, $i$, 2MASS $J, H, K$, WISE $W1$, $W2$, and Spitzer [3.6] and [4.5] apparent magnitudes listed in Table 1.

Our SED fit had six free parameters: the effective temperature and surface gravity of CWW 89A, the effective temperature and surface gravity of CWW 89B, the distance to CWW 89, and the amount of visual extinction, $A_V$, to CWW 89. We imposed Gaussian priors on the effective temperature and surface gravity of CWW 89A based on the spectroscopic properties listed in Table \ref{tab:stellarprops}, and on the measured distance and $A_V$ to R147 \citep{curtis2013}. We did not impose priors on the effective temperature or surface gravity of CWW 89B, other than initializing the values of both at 3700\,K and $\log g=4.8$, respectively. We fixed the metallicity of both stars to be zero. We also included a Gaussian prior on the $K$-band brightness ratio of the two stars, based on the measured $\Delta K$ from the AO images.

To determine the SEDs for both stars we used the Castelli-Kurucz model atmospheres \citep{ck2004} to compute a grid of surface luminosity magnitudes that matched our catalog apparent magnitudes in ($\teff$, $\log g$) space. Since the Castelli-Kurucz models step by 250\,K in $\teff$ and 0.5 dex in $\log g$,, we used a cubic spline interpolation to estimate model magnitudes in between the points provided by the Castelli-Kurucz atmospheres. We then scaled these interpolated surface magnitudes by $(R_*/d)^2$, where $R_*$ is the radius of either CWW 89A or CWW 89B, and $d$ is the distance to the system. We used the Torres mass-radius relations for main sequence stars \citep{torres2010} to estimate the radii of both stars based on their effective temperatures and surface gravities. Finally, we applied a simple $R = 3.1$ extinction law scaled from the value of $A_V$ to determine the extincted bolometric flux of the combined SED model.

\begin{deluxetable}{lcc}
\tablecaption{Properties of the CWW 89 Stellar System}
\tablehead{\colhead{~~~Parameter} & \colhead{Value} & \colhead{Ref.}}
\startdata
\sidehead{CWW-89 System}
\hline
~~~Age (Gyr) & $3.00\pm0.25$ & C13\\
~~~Distance (pc) & $307.6\pm4.6$ & Gaia DR2\\
~~~$A_V$ (mag) & $0.25\pm0.05$ & C13\\
\hline
\sidehead{CWW-89A}
\hline
~~~$T_\mathrm{eff}$ (K) & $5715\pm80$ & C18\\
~~~$\log(g)$ & $4.48\pm0.05$ & C18\\
~~~$\feh$ & $0.04\pm0.03$ & C18\\
~~~v$\sin(i)$ (km s$^{-1}$) & $4.5\pm0.5$ & C18\\
~~~$M_*$ (\msun) & $0.99\pm0.05$ & C18\\
~~~$R_*$ (\rsun) & $1.01\pm0.04$ & C18\\
\hline
\sidehead{CWW-89B}
\hline
~~~$T_\mathrm{eff}$ (K) & $3800\pm275$ & This work\\
~~~$\log(g)$ & $4.7\pm0.3$ & This work\\
~~~$\feh$ & $\equiv0.0$ & This work\\
~~~$M_*$ (\msun) & $0.52\pm0.06$ & This work\\
~~~$R_*$ (\rsun) & $0.52\pm0.06$ & This work
\enddata
\tablecomments{C13 = \cite{curtis2013}, C18 = \cite{curtis2018}}
\vskip -0.25in
\label{tab:stellarprops}
\end{deluxetable} 

We used the \emph{emcee} python package to conduct a Markov Chain Monte Carlo (MCMC) optimization of this likelihood and to fit the SED model. We used twenty walkers and used a 300 step burn-in and 2,000 step production run. We initialized the walkers in a random Gaussian ball about the initial starting location in parameter space. At the end of the production run the Gelman-Rubin statistic for each parameter was less than 1.1, and a visual inspection of the covariance corner plot showed a set of well behaved posteriors, so we judged the MCMC process to have converged.

We find that CWW 89B is most likely an early M-dwarf with an effective temperature of $3800\pm275$\,K and $\log g = 4.7\pm0.3$ (cgs). We list the complete set of properties we determine for CWW 89B in Table \ref{tab:stellarprops}. At the wavelengths of our Spitzer observations CWW 89B contributes $14.6\pm0.8$\,\% of the total system light at \three and $14.1\pm1.0$\,\% at 4.5\um\ (Figure \ref{sed}). We used this dilution measurement to correct the secondary eclipse depths after our fitting process, and to determine the undiluted apparent magnitude of CWW 89A.

\subsection{CWW 89's Membership in Ruprecht-147}

\cite{curtis2013} listed CWW 89 as a ``probable'' member of R147, based on a combination of radial velocity, proper motion, and color measurements. While they found that measured proper motion and colors of CWW 89 were consistent with cluster membership, CWW 89 showed a systemic radial velocity of $47\pm7$ km s$^{-1}$, which was nominally higher than the membership cut-off at 43 km s$^{-1}$. By resolving the reflex motion of CWW 89A due to CWW 89Ab, \cite{nowak2017} later refined the system's systemic velocity to be $45.8\pm0.1$ km s$^{-1}$. This is approximately, and significantly, above \cite{curtis2013}'s 43 km s$^{-1}$ systemic velocity cut-off.

However, orbital motion due to the newly identified binary companion CWW 89B can explain the high system velocity seen for CWW 89A. As a basic example, if the binary orbit is circular, edge-on to our line of sight, and has a semi-major axis of 24.9 AU (the projected separation of A and B), then the semi-amplitude of the binary's radial velocity orbit is 4 km s$^{-1}$ -- which could then explain the 2.8 km s$^{-1}$ difference between CWW 89A's  and R147's systemic velocities. Needless to say, in reality we do not have any constraint on the geometry of the binary orbit, but many of the possible geometries (e.g., increasing the eccentricity in the example above) yield binary orbital velocities capable of accounting for the apparent velocity discrepancy between CWW 89A and the cluster. At the very least, then, the measured systemic velocity of CWW 89A is a ``draw,'' and does not provide strong evidence for, or against, membership in R147.

This leaves us with the parallax, proper motion, and colors of CWW 89A to determine its cluster membership. As mentioned, \cite{curtis2013} found the latter consistent with membership, which is still the case. As for CWW 89A's parallax and proper motion, the recent Gaia DR2 results \citep{gaiadr2} are completely consistent with CWW 89A being a cluster member: both the star's distance and its position in velocity space are close to the center of the membership diagrams used in \cite{curtis2013}.

Therefore, since the Gaia DR2 parallax and proper motion are indicative of cluster membership, and the apparent RV discrepancy can be explained by the presence of CWW 89B, we are confident that the CWW 89 system is a member of R147.

\section{Lightcurve Modeling, Fitting Process, and Results}

Our raw extracted photometry in both bands displayed the usual position dependent systematics present in Spitzer \textsc{irac} photometry due to intra-pixel sensitivity variations (Figure \ref{rawplot}). To account for these effects, we used the BiLinearly-Interpolated Subpixel Sensitivity (BLISS) mapping technique to simultaneously fit a sub-pixel sensitivity map along with an astrophysical eclipse model. 

In addition to the usual systematics, the \three photometry also showed a substantial (4\,mmag) bump about half way through the predicted time of the secondary eclipse. As we discuss in Section 3.3, we believe that this bump was caused by a coincidental stellar flare, and so we also included a stellar flare model for the \three photometry. 

\begin{deluxetable}{lcl}[b!]
\tablecaption{Prior Values for CWW 89Ab's Properties\\ From \cite{curtis2018}}
\tablehead{\colhead{Parameter} & \colhead{Units} & \colhead{Value}}
\startdata
$T_C$\dotfill &Transit time (\bjdtdb)\dotfill & $2457341.03701\pm0.00012$\\
$P$\dotfill &Orbital period (days)\dotfill & $5.292601\pm0.000025$\\
$\sqrt{e}\cos{\omega}$\tablenotemark{a}\dotfill & \dotfill & $0.4228\pm0.0025$\\
$\sqrt{e}\sin{\omega}$\tablenotemark{a}\dotfill & \dotfill & $-0.1062\pm0.0006$\\
$\cos{i}$\dotfill &Cosine of inclination\dotfill & $0.0021\pm0.0016$\\
$R_{P}/R_{*}$\dotfill &Radius ratio\dotfill & $0.09350\pm0.00072$\\
$a/R_{*}$\dotfill &Scaled semimajor axis\dotfill & $13.88\pm0.40$\\
\hline
$M_{BD}$\tablenotemark{b}\dotfill &Brown dwarf mass (\mj)\dotfill & $36.5\pm0.1$\\
$R_{BD}$\tablenotemark{b}\dotfill &Brown dwarf radius (\rj)\dotfill & $0.937\pm0.042$\\
$\log(g)_{BD}$\tablenotemark{b}\dotfill &Brown dwarf gravity (cgs)\dotfill & $5.01\pm0.04$
\enddata
\tablenotetext{a}{Calculated from the measured $e$ and $\omega$ in \cite{curtis2018}.}
\tablenotetext{b}{Not a fitting parameter, but provided for reference.}
\vskip -0.35in
\label{tab:priors}
\end{deluxetable}

Our complete model for the observed flux was thus 
\begin{equation}\label{eq:3010}
F_{obs,3.6} = (E[\Theta_{ecl},t] + Fl[\Theta_{fl},t])\,B(x,y,t)\,R(r_1,t)
\end{equation}
at \three and 
\begin{equation}\label{eq:3020}
F_{obs,4.5} = E(\Theta_{ecl},t)\,\,B(x,y,t)\,R(r_1,t)
\end{equation}
at 4.5\um. The astrophysical portions of the above equations are $E(\Theta_{ecl},t)$, the astrophysical eclipse model based on the physical system parameters $\Theta_{ecl}$ and time $t$, and the \three flare model $Fl(\Theta_{fl},t)$, also based on a set of flare parameters $\Theta_{fl}$ and time. $B(x,y,t)$ is the sensitivity of the BLISS map given the stellar $x$ and $y$ pixel position and time, and $R(r_1,t)$ is a background linear ramp present to account for a long term trend in both channels.

\subsection{Secondary Eclipse Model, Parameters, and Priors}

For the secondary eclipse model $E(\Theta_{ecl},t)$ we used the \textsc{batman} python package \citep{kreidberg2015} to generate an eclipse lightcurve. \textsc{batman} is an implementation of a \cite{mandel2002} lightcurve model. We calculated the lightcurve models using the observed transit cetner time ($T_C$), the orbital period (as $\log[P]$), $\sqrt{e}\cos{\omega}$, $\sqrt{e}\sin{\omega}$, the cosine of the orbital inclination ($\cos i$), the radius of the planet in stellar radii ($R_P/R_*$), the semi-major axis in units of the stellar radii (as $\log[a/R_*]$), and the depth of the secondary eclipse itself ($\delta$). This gave us eight physical parameters which set our secondary eclipse model:
\begin{eqnarray}\label{eq:3110}
\Theta_{ecl} = && (T_C,\log P,\sqrt{e}\cos\omega,\sqrt{e}\sin\omega,\cos i, \\ \nonumber
&& R_P/R_*,\log a/R_*,\delta).
\end{eqnarray}
Note that we use the observed transit center time, rather than a time of secondary eclipse, in our eclipse model because no secondary eclipse for this system has been observed. We therefore calculated the expected secondary eclipse times based on $T_C$, $\log P$, $\sqrt{e}\cos{\omega}$, and $\sqrt{e}\sin{\omega}$.

\begin{figure}[t]
\vskip 0.00in
\includegraphics[width=1.0\linewidth,clip]{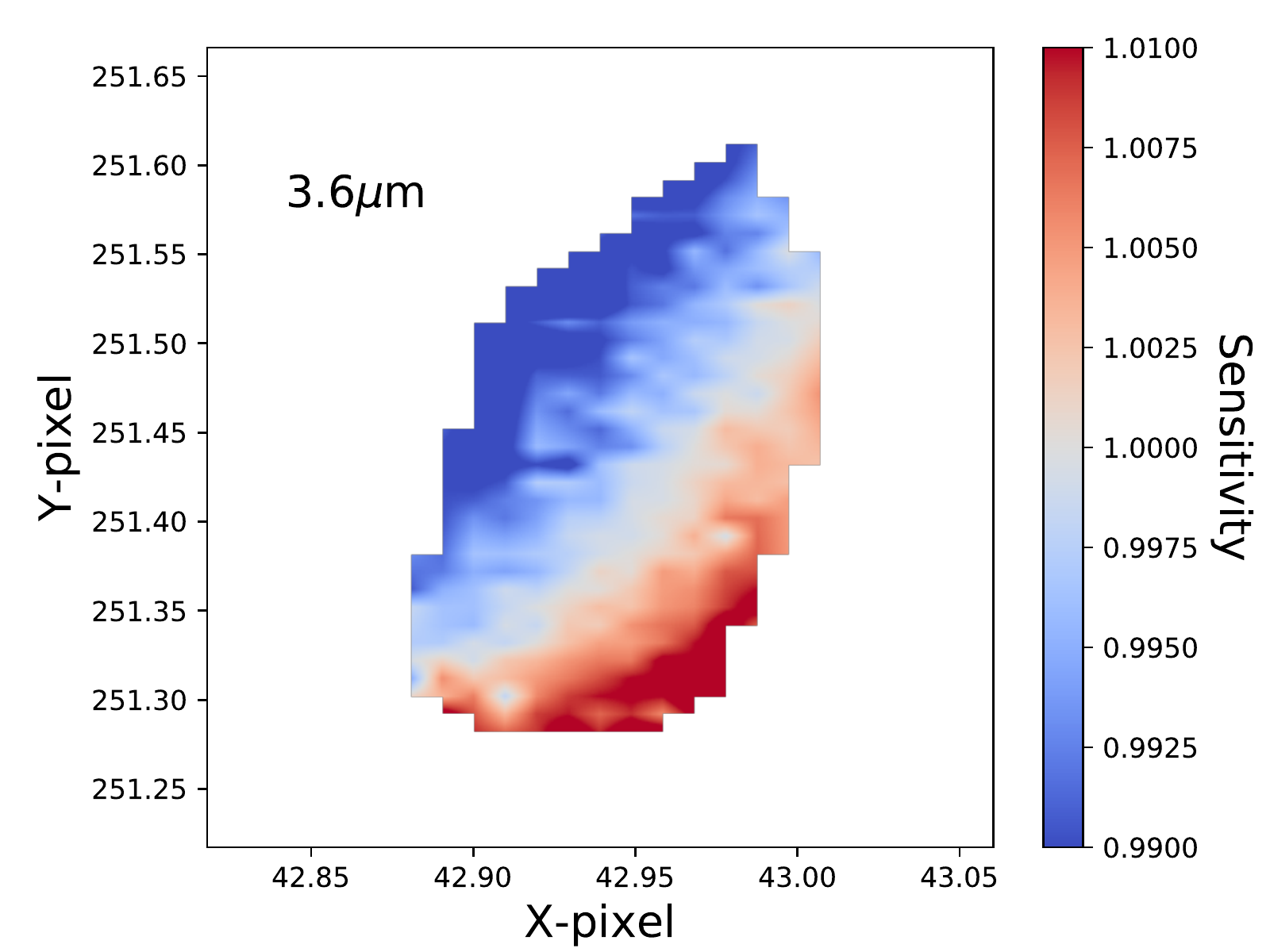}
\vskip -0.0in
\caption{BLISS sensitivity map for our \three observations, as described in Section 3.2.}
\label{blissgrid36}
\end{figure}

All of the secondary eclipse model parameters except for the secondary eclipse depth were previously measured using the K2 discovery photometry. We therefore imposed Gaussian priors on these seven parameters based on the measurements in \cite{curtis2018}, and listed in Table \ref{tab:priors}. We do not impose any explicit prior on the secondary eclipse depth, which implicitly imposes a uniform prior.

\subsection{BLISS Model and Ramp}

As noted previously, the raw Spitzer photometry in both bands showed correlations with the pixel position of the stellar centroid in each image. This is a typical feature of \textsc{irac} photometry during Spitzer's extended warm mission that arises due to inter- and intrapixel sensitivity variations in the \textsc{irac} detectors, and many different methods have been used to treat these correlations. For these observations, we chose to use \cite{stevenson2012}'s BLISS mapping technique.

\begin{figure}[t]
\vskip 0.00in
\includegraphics[width=1.0\linewidth,clip]{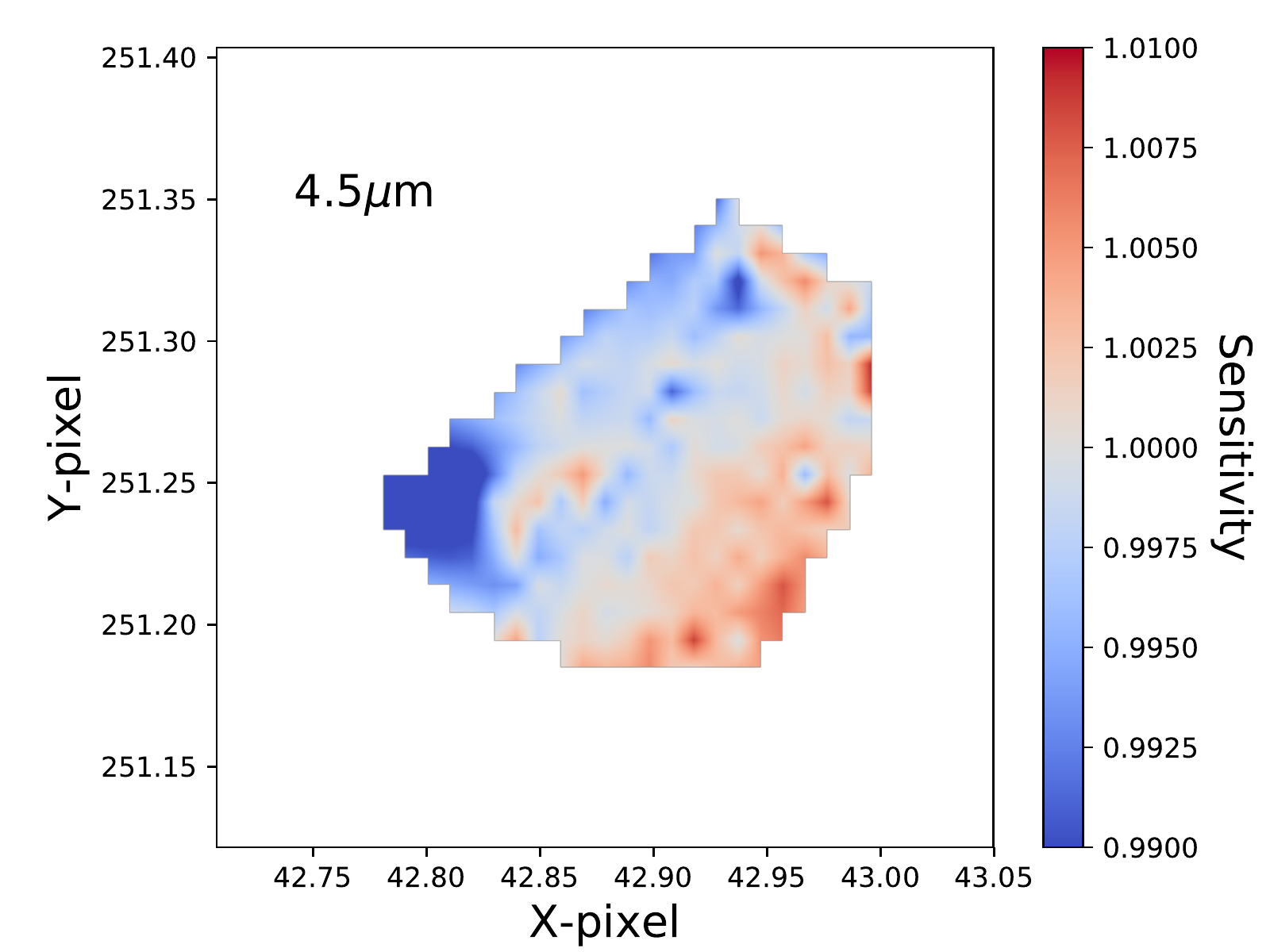}
\vskip -0.0in
\caption{BLISS sensitivity map for the \four observations, as described in Section 3.2.}
\label{blissgrid45}
\end{figure}

BLISS mapping attempts to fit the underlying position-dependent sensitivity of the \textsc{irac} detectors simultaneously with the astrophysical signal. To do so, BLISS mapping begins by taking the residuals in the raw photometry to the proposed astrophysical signal (here $E(\Theta_{ecl},t)$), and assuming that these residuals are predominantly due to detector sensitivity variations. Using the measured x- and y-pixel positions of the stellar centroid, BLISS mapping then models the detector sensitivity by constructing a bilinear interpolation of the photometric residuals as a function of x- and y-pixel position. Thus for a given secondary eclipse model
\begin{equation}\label{eq:3210}
B(x,y,t) = \frac{F_{obs}}{E[\Theta_{ecl},t]\,R(r_1,t)}.
\end{equation}
Following \cite{stevenson2012}, we also included a linear ramp term, $R(r_1,t)$ in our systematic noise model. This had the form
\begin{equation}\label{eq:3220}
R(r_1,t) = r_1\,(t-\tilde{t}) + 1,
\end{equation}
where $\tilde{t}$ was the median observation time in a given channel and the slope $r_1$ was a free parameter. Figures \ref{blissgrid36} and \ref{blissgrid45} show the BLISS maps that we determined as a part of the final fitting process.

\subsection{\three Flare Model}

In addition to the predicted secondary eclipse, our \three photometry showed a noticeable upward ``bump'' shortly after mid-eclipse, at approximately \bjdtdb\ 2457751.85. Our initial assumption was that this was some untreated systematic noise in the \three photometry, but closer examination of CWW 89's measured pixel positions showed no corresponding jump in either the x- or y- position (bottom left two panels of Figure \ref{rawplot}). Similarly, the FWHM of the stellar PSF did not suddenly change at this time, nor was there anything visible in the \textsc{irac} images themselves to explain the bump in the \three photometry. This left us to consider possible astrophysical causes.

\begin{figure*}[t]
\vskip 0.00in
\includegraphics[width=1.0\linewidth,clip]{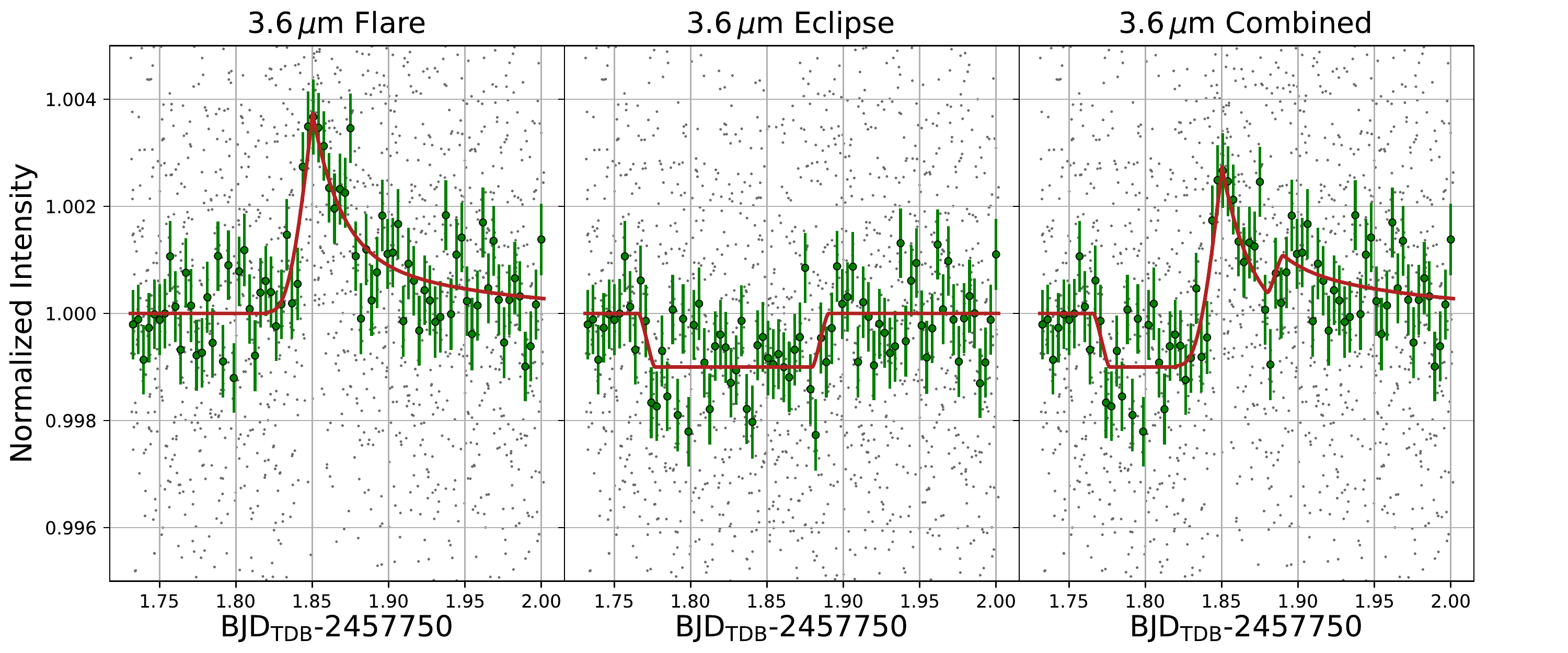}
\vskip -0.0in
\caption{The detrended \three photometry showed both the secondary eclipse of CWW 89Ab, and an apparent stellar flare at approximately \bjdtdb\ 2457751.85 (rightmost panel). We modeled the lightcurve as the additive combination of the flare model from \cite{davenport2014} and an eclipse model. In the leftmost panel, we show the detrended \three photometry with the secondary eclipse model subtracted to illustrate the bestfit flare model. In the middle panel, we then show the detrended \three photometry with the flare model subtracted to show the bestfit secondary eclipse. Note that we fit both the secondary eclipse and flare simultaneously (along with the \four photometry). In all panels, the small grey points are the unbinned \three photometry, and the green points with uncertainties are the data binned into 5 minute intervals.}
\label{nice36}
\end{figure*}

\begin{figure}
\vskip 0.00in
\includegraphics[width=1.1\linewidth,clip]{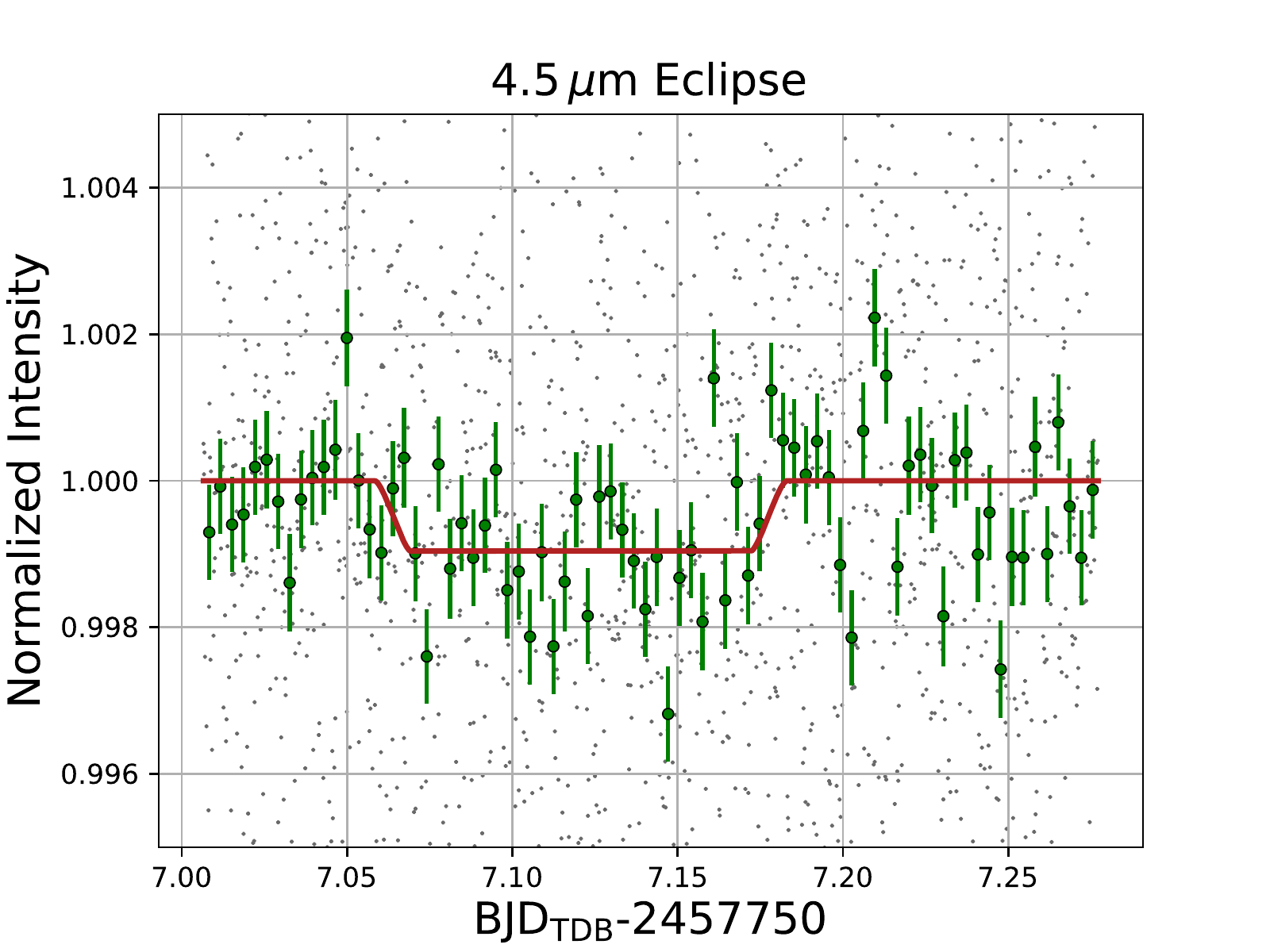}
\vskip -0.0in
\caption{The bestfit \four secondary eclipse, which we fit simultaneously alongside the \three data. The small grey points are the unbinned \three photometry, and the green points with uncertainties are the data binned into 5 minute intervals.}
\label{nice45}
\end{figure}

The easiest way to cause an upward jump in our photometry would be for one of the stars in the CWW 89 system to have had a stellar flare. To test this hypothesis, we took the stellar flare model from \cite{davenport2014} and included it in our model for the \three photometry. The \cite{davenport2014} model is an empirical description of flare morphology based on combining more than 6100 flares from the M4V star GJ 1243 that were observed by the Kepler spacecraft in the red-optical. The \cite{davenport2014} flare model has three free parameters,
\begin{equation}\label{eq:3310}
\Theta_{fl} = (A_{fl},T_{peak},T_{1/2}),
\end{equation}
which are the flare amplitude $A_{fl}$, the peak time $T_{peak}$, and a characteristic timescale $T_{1/2}$. This single characteristic timescale sets both the rise and decay lengths of the flare. The functional forms of the rise and decay are fixed to be a quadratic polynomial rise and a double exponential decay, and the exact functional form that we used for the flare model, $Fl(\Theta_{fl},t)$, is given by Equations 1 and 4 in \cite{davenport2014}.

\subsection{Joint Fitting Process and Results}

We fit both the \three and \four observations simultaneously using a single set of orbital parameters and a single value of $R_p/R_*$, but individual secondary eclipse depths, BLISS maps, and ramp functions for each channel. The parameters of the \three flare model were unique to the \three observations. As described in Section 3.1, we imposed Gaussian priors on the non-eclipse depth physical properties of the system based on the measurements in \cite{curtis2018} and as listed in Table \ref{tab:priors}. Note too that we fit the observed photometry without correcting for the presence of the unresolved companion star CWW 89B -- we applied the dilution correction determined in Section 2.2 after this fitting process.

\begin{deluxetable*}{lcl}
\tabletypesize{\footnotesize}
\tablecaption{Median Values and 68\% Confidence Intervals \\ for the Joint \three and \four Fit}
\tablehead{\colhead{~~~Parameter} & \colhead{Units} & \colhead{Value}}
\startdata               
\sidehead{Secondary Eclipse Model Parameters:}
~~~$T_C$\dotfill &Transit time (\bjdtdb)\dotfill & $2457341.03700\pm0.00012$\\
~~~$\log(P)$\dotfill &Log orbital period (days)\dotfill & $0.7236688\pm2.0\times10^{-6}$\\
~~~$\sqrt{e}\cos{\omega}$\dotfill & \dotfill & $0.4213\pm0.0011$\\
~~~$\sqrt{e}\sin{\omega}$\dotfill & \dotfill & $-0.1061\pm0.0006$\\                     
~~~$\cos{i}$\dotfill & Cosine of inclination\dotfill & $0.0023\pm0.0014$\\
~~~$R_{P}/R_{*}$\dotfill &Radius of planet in stellar radii\dotfill & $0.09347\pm0.00070$\\
~~~$\log(a/R_{*})$\dotfill &Log semi-major axis in stellar radii\dotfill & $1.142\pm0.011$\\
~~~$\delta_{3.6,dil}$\dotfill &Diluted \three eclipse depth (ppm)\dotfill & $1001\pm199$\\
~~~$\delta_{4.5,dil}$\dotfill &Diluted \four eclipse depth (ppm)\dotfill & $959\pm209$\\
\sidehead{\three Flare Model Parameters:}
~~~$A_{fl}$\dotfill & Amplitude (ppm)\dotfill & $3731\pm471$\\
~~~$T_{peak}$\dotfill &Peak time (\bjdtdb)\dotfill & $2457751.8501\pm0.0017$\\
~~~$T_{1/2}$\dotfill &Characteristic time (days)\dotfill & $0.0292\pm0.007$\\
\sidehead{BLISS Model Parameters:}
~~~$r_{1,3.6}$\dotfill &\three linear ramp\dotfill & $0.0037\pm0.0016$\\
~~~$r_{1,4.5}$\dotfill &\four linear ramp\dotfill & $0.0000\pm0.0035$\\
\sidehead{Derived Parameters:}
~~~$T_S$\dotfill &Secondary eclipse time (\bjdtdb)\dotfill & $2457751.8281\pm0.0028$\\
~~~$P$\dotfill &Orbital period (days)\dotfill & $5.292597\pm0.000024$\\
~~~$a/R_{*}$\dotfill &Semi-major axis in stellar radii\dotfill & $13.87\pm0.36$\\
~~~$i$\dotfill &Inclination (degrees)\dotfill & $89.87\pm0.08$\\
~~~$b$\dotfill &Impact Parameter\dotfill & $0.03\pm0.02$\\
~~~$\tau$\dotfill &Ingress/egress duration (days)\dotfill & $0.01173\pm0.00032$\\
~~~$T_{14}$\dotfill &Total duration (days)\dotfill & $0.1368\pm0.0035$\\\
~~~$e$\dotfill &Orbital Eccentricity\dotfill & $0.1888\pm0.0010$\\
~~~$\omega$\dotfill &Argument of periastron (degrees)\dotfill & $345.9\pm0.1$\\
\sidehead{Dilution Corrected Eclipse Depths and \textsc{irac} Magnitudes:}
~~~$\delta_{3.6}$\dotfill &\three eclipse depth (ppm)\dotfill & $1147\pm213$\\
~~~$\delta_{4.5}$\dotfill &\four eclipse depth (ppm)\dotfill & $1097\pm225$\\
~~~$[3.6]$\dotfill &\three apparent mag. (dayside)\dotfill & $17.98\pm0.20$\\
~~~$[4.5]$\dotfill &\four apparent mag.  (dayside)\dotfill & $18.05\pm0.22$\\
~~~$\mathrm{Abs.}\ [3.6]$\dotfill &\three absolute mag. (dayside)\dotfill & $10.54\pm0.22$\\
~~~$\mathrm{Abs.}\ [4.5]$\dotfill &\four absolute mag.  (dayside)\dotfill & $10.61\pm0.23$
\enddata
\label{tab:results}
\end{deluxetable*}

We began by conducting an initial Nelder-Mead maximization to identify a beginning best fit to the observations. We then took this initial best fit and used an MCMC process to determine uncertainties on our parameter measurements, and to verify that our fitting had correctly settled on the global maximum likelihood. We used the \emph{emcee} python package \citep{dfm2013} to run our MCMC fit. For the MCMC fitting we used 40 walkers, which we initialized in a random Gaussian ball centered on the results of the Nelder-Mead fit. We then ran the MCMC process forward using a 1,000 step burn-in, followed by a 15,000 step production run.

We used the Gelman-Rubin (GR) test statistic to determine if the resulting MCMC chains had converged. We required the GR statistic for each parameter to be below 1.01 for successful convergence. We do note that since \emph{emcee} uses an affine invariant sampler, the GR statistic is not entirely accurate in our case, since Gelman-Rubin assumes that each of the MCMC chains are independent of each other. As an alternate metric we also calculated the autocorrelation length of each parameter's MCMC chains, which ranged from about 200 to 270. Since even the maximum autocorrelation length was less than 1/50 the length of an individual walker's chain, we judged this to also be strong evidence for convergence.

\begin{figure*}[t]
\vskip 0.00in
\includegraphics[width=1.0\linewidth,clip]{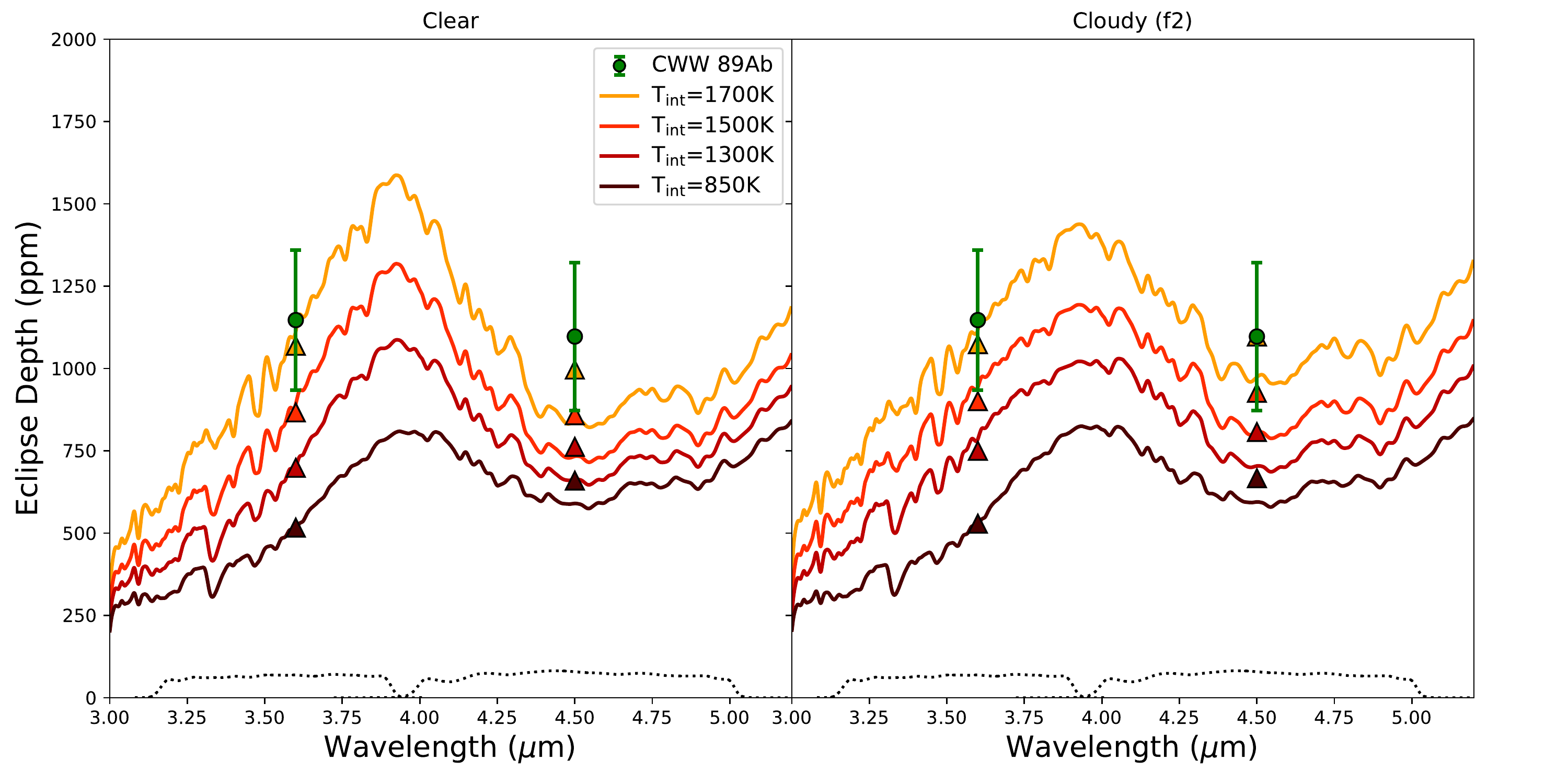}
\vskip -0.0in
\caption{Based on the measured secondary eclipse depths (green points) the clear (left side) and cloudy (right side) irradiated brown dwarf atmosphere models described in Section 3.5 predict that CWW 89Ab has an internal luminosity of $T_{\rm int}=1700\pm130$\,K, or $\log(\mathrm{L_{bol}/\mathrm{L}_\odot})=-4.19\pm0.14$ -- after accounting for the incoming stellar irradiation. This is $9.3\,\sigma$ more luminous than predicted by brown dwarf evolutionary models, which expect $T_{\rm int}=850$\,K, or $\log(\mathrm{L_{bol}/\mathrm{L}_\odot})=-5.4$.}
\label{morley}
\end{figure*}

Table 4 lists the best fit from the MCMC fitting, along with uncertainties. After applying the dilution correction we determined in Section 2.2, we find undiluted secondary eclipse depths of $1147\pm213$\,ppm at \three and $1097\pm225$\,ppm at \fouralt. These secondary eclipse depths mean that the dayside of CWW 89Ab has \three and \four apparent magnitudes of $[3.6]=17.98\pm0.20$ and $[4.5]=18.05\pm0.22$. Using the measured distance to Ruprecht-147 \citep{curtis2013}, this places the dayside of CWW 89Ab squarely in the middle of the field brown dwarf sequence on an IRAC color-magnitude diagram \citep{dupuy2012}, and is consistent with a dayside spectral type of approximately L4.

Figures \ref{nice36} and \ref{nice45} show the detrended secondary eclipse photometry at \three and \fouralt, respectively. In Figure \ref{nice36} we have decomposed the photometry into just the flare component (left panel), just the secondary eclipse component (middle panel), and the actual, combined, photometry (right panel), to help visualize each individual model. 

The standard deviation of the residuals to our joint fit was 3,325\,ppm at \three and 3,863\,ppm at 4.5\um. We tested the Gaussianity of the residuals using an Anderson-Darling test and calculated a test statistic of 0.20 for the \three residuals and 0.26 for the \four residuals, which indicates that both sets of residuals are consistent with a Gaussian distribution. For reference, the test statistic would need to be greater then unity for the residuals to be significantly non-Gaussian.

\subsection{Atmospheric Modeling}

We modeled the thermal emission spectrum from CWW 89Ab's atmosphere using methods described in more detail in \citet{fortney2008} and \citet{morley2015}. We calculated pressure--temperature profiles and resulting thermal emission spectra assuming radiative--convective equilibrium and chemical equilibrium, including both internal heat ($\mathrm{T}_\mathrm{int}$) and the incident flux from the host star. All models had solar metallicity and solar C/O ratio. We included clouds made of silicates, iron, and corundum in some models using a modified version of the \citet{ackerman2001} cloud model. We varied the internal heat ($\mathrm{T}_\mathrm{int}$) from 850 K (the interior temperature predicted for an isolated brown dwarf with the age and mass of CWW 89Ab) to 1700 K. The resulting models are shown in Figure \ref{morley} along with the measured secondary eclipse depths.

Based on our modeling we find that the internal luminosity of CWW 89Ab is $T_{\rm int}=1700\pm130$\,K, or $\log(\mathrm{L_{bol}/\mathrm{L}_\odot})=-4.19\pm0.14$. As we discuss below, this is significantly more luminous than predicted by brown dwarf evolutionary models, which expect $T_{\rm int}=850$\,K, or $\log(\mathrm{L_{bol}/\mathrm{L}_\odot})=-5.4$.

The models that we adopt, and that are shown in Figure \ref{morley}, all assume no heat redistribution from the day- to nightside of CWW 89Ab, such that all the incoming stellar energy is re-radiated by the substellar hemisphere. We note that this assumed low heat redistribution efficiency is counter to trends seen in hot Jupiters, which show near complete heat redistribution at these temperatures \citep{komacek2017}. However, any increase the atmospheric heat redistribution efficiency would only serve to exacerbate the over-luminosity described above, since the dayside irradiation-only temperature would be corresponding lower. For example, if we were to assume complete heat redistribution in CWW 89Ab, then we would find that $T_{\rm int}\approx2300$\,K. We have therefore conservatively assumed the case of no heat redistribution from day to night, rather than further increasing the interior temperature of CWW 89Ab.

Given the deeper than predicted secondary eclipse depths, we also note that we did not include a possible reflected light signal in our models. This is because the expected secondary eclipse depth from reflected light alone is $\delta_\mathrm{refl}=A_g\,(Rp/a)^2=A_g\,(45\,\mathrm{ppm})$, where $A_g$ is the geometric albedo of the dayside at a particular wavelength. The entire range of the reflection signal is thus four times smaller than the uncertainties on the secondary eclipse depths. In addition, in the unlikely, extreme, case of $A_g\sim1$ in the IRAC bandpasses the necessary optically thick reflecting clouds would also totally obscure the internal thermal emission -- which is impossible given these observations.

\section{The High Interior Luminosity of CWW 89Ab, and Possible Explanations}

CWW 89Ab's inferred interior luminosity of $\log(\mathrm{L_{bol}/\mathrm{L}_\odot})=-4.19\pm0.14$ is starkly higher than predictions from evolutionary models (Figure \ref{agelum}). The non-irradiated \cite{sm08} f2 evolutionary model (hereafter SM08) at 3 Gyr predict that CWW 89Ab's interior luminosity should be $\log(\mathrm{L_{bol}/\mathrm{L}_\odot})=-5.4$. This is 16 times, or 1.2 dex, lower than observed. A set of irradiated evolutionary models based on \cite{burrows1997, burrows2001} and \cite{spiegel2010, spiegel2012, spiegel2013} (hereafter the ``Burrows'' models) that match the irradiation received by CWW 89Ab also predict an interior luminosity of $\log(\mathrm{L_{bol}/\mathrm{L}_\odot})=-5.4$. In both cases this is a difference of $9.3\,\sigma$ to the observations.

\subsection{Unworkable Explanations}

\begin{figure}[t]
\vskip -0.0in
\includegraphics[width=1.1\linewidth]{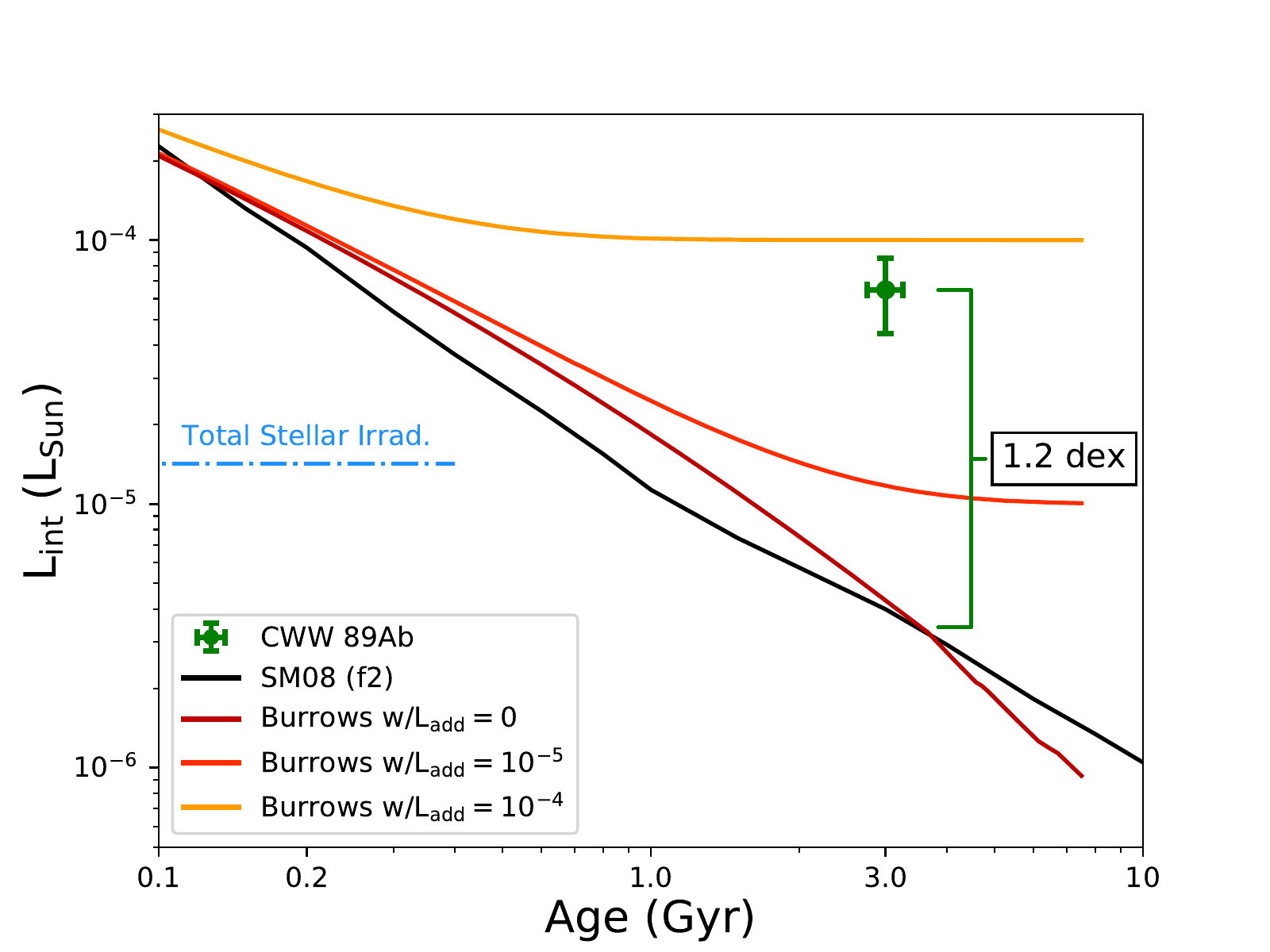}
\vskip -0.0in
\caption{The measured age and luminosity of CWW 89Ab (green point) plotted on top of models from SM08 and Burrows (see the beginning of Section 4 for references). The observed luminosity is 1.2 dex (16 times) larger than either of the models. Note that though we have marked the total amount of stellar irradiation received by CWW 89Ab mid-way up the left axis, this has already been accounted for in our luminosity determination: the observed over-luminosity is in addition to the incoming stellar energy. Though an additional interior luminosity would explain the observations (i.e., the orange line), it would inflate the radius of CWW 89Ab far beyond what is observed (Figure \ref{massradius}).}
\vskip -0.0in
\label{agelum}
\end{figure}

1) \emph{Inaccurate age determination:} At first glance, the most straightforward explanation for the observed over-luminosity in CWW 89Ab is that either the CWW 89 system, or CWW 89Ab itself, are younger than the cluster age of 3.0 Gyr. This could occur if \cite{curtis2013} misidentified CWW 89 as a cluster member, though the Gaia DR2  \citep{gaiadr2} parallax and proper motion for the system strongly support CWW 89's cluster membership (Section 2.3). Alternately, CWW 89Ab may have formed later and independently from the rest of stellar system and was captured in a three-body interaction with the binary. In these cases, either the stellar system or the brown dwarf would need to be approximately 0.25 Gyr old for the internal luminosity predicted by either the SM08 or Burrows models to match the observed luminosity of CWW 89Ab (Figure \ref{agelum}).

Leaving aside the improbability of the binary capturing a free-floating brown dwarf \citep[e.g.,][]{bodenheimer2011}, an inaccurate age determination cannot be causing the over-luminosity for two reasons. First,  the observed rotation period for CWW 89A of 12.66 days \citep{curtis2016, nowak2017} is too slow for the system to be only 0.25 Gyr old. As a Sun-like star the rotation period of CWW 89A should be well determined by gyrochronology relations, and observations of the 193 Myr old cluster M34 show that CWW 89A's rotation period should be, at most, about 4 days \citep{james2010} if it were 0.25 Gyr old. At 12.66 days, gyrochronology instead predicts that CWW 89A should be at least 1 Gyr old \citep{meibom2011}, and the theoretical rotation period at the 3 Gyr age of the cluster is $20\pm0.5$ days (J. van Saders, private communication), which is roughly the same as the observed 18 day rotation periods of Solar-like stars seen in the 2.5 Gyr cluster NGC 6819 \citep{meibom2015}. Though this would seem to indicate that CWW 89A is instead 1 Gyr old, given the relatively high mass of CWW 89Ab and its eccentric orbit, it seems extremely likely that tidal interactions have spun-up CWW 89A to the observed 12.66 day period (see Section 5).

\begin{figure}[t]
\vskip -0.0in
\includegraphics[width=1.1\linewidth,clip]{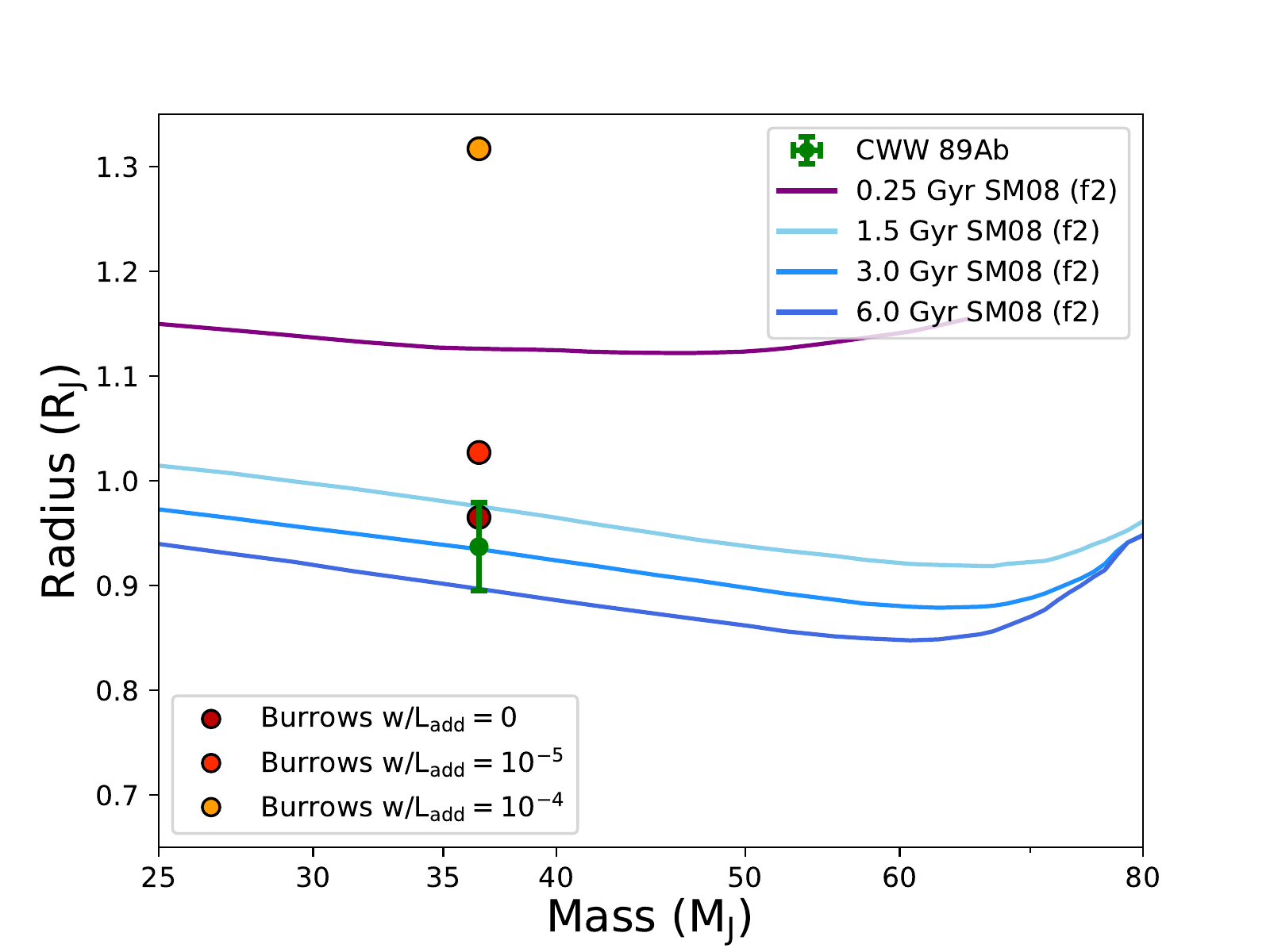}
\vskip -0.0in
\caption{The measured mass and radius of CWW 89Ab (green point) plotted on top of models from SM08 and Burrows (see the beginning of Section 4 for references). Both sets of models correctly predict the measured mass and radius of CWW 89Ab for an age of 3.0 Gyr. If we attempt to explain the observed over-luminosity by assuming CWW 89Ab is significantly younger than this (purple line), or that there is some source of additional interior luminosity (orange point), the models do not reproduce the observed radius. Compare to Figure \ref{agelum}.}
\label{massradius}
\end{figure}

Secondly, and more dammingly, if CWW 89Ab were 0.25 Gyr old the observed discrepancy would simply shift from an over-luminosity to an under-inflated radius. Recall that at 3.0 Gyr and 36.5\mj\ evolutionary models nearly perfectly predict CWW 89Ab's observed radius of 0.94\rj. If CWW 89Ab were, instead, 0.25 Gyr, the SM08 models instead predict that the radius would be 1.15\rj, which is 4.5\,$\sigma$ larger than observed (Figure \ref{massradius}). Similarly, if one believes that the stellar rotation period has been unaffected by tides from CWW 89Ab (though see Section 5), and the system age is roughly 1 Gyr from gyrochronology \citep{meibom2011}, the model predictions still do not agree with the observations. The ``best case'' scenario at 1 Gyr is the Burrows $\log(\mathrm{L_{add}/\mathrm{L}_\odot})=-5$ model, which is excluded by the combined luminosity and radius measurements at 5.0\,$\sigma$.
\\ \newline
2) \emph{Stellar-heating driven luminosity:} This group of scenarios encompasses the set of explanations that depend upon the incoming stellar irradiation to drive the observed over-luminosity. Before discussing these, we reiterate that the atmospheric models we describe in Section 4 and in Figure \ref{morley} \emph{already account for stellar heating of the dayside,} and that the over-luminosity we observe is in addition to the heating we expect to see from stellar irradiation. It thus follows that most stellar-heating driven explanations for CWW 89Ab's over-luminosity are double-counting the incoming stellar irradiation.

Nevertheless, some heating mechanisms, like Ohmic dissipation \citep{ohmic2010}, may be able to simultaneously heat the observed photosphere and the brown dwarf interior. This is especially true for CWW 89Ab, since the large surface magnetic fields of 10G to 100G observed in isolated brown dwarfs \citep{berger2006} imply that Ohmic dissipation will be particularly efficient. \cite{menou2012} estimates that in the case of strong magnetic fields Ohmic dissipation would be able to transfer approximately 10\% of the incoming stellar irradiation into the interior. Since CWW 89Ab intercepts $5.47\times10^{21}$\,W, or $\log(\mathrm{L/\mathrm{L}_\odot})=-4.85$, of stellar energy from CWW 89A assuming an albedo of zero, ``peak'' Ohmic dissipation would thus only provide an interior luminosity of $\log(\mathrm{L_{ohm}/\mathrm{L}_\odot})=-5.85$. This is much lower than the observed $\log(\mathrm{L_{bol}/\mathrm{L}_\odot})=-4.19\pm0.14$.

Indeed, no matter what method is hypothesized to cause the incoming stellar irradiation to heat the atmosphere, it will be impossible for the atmosphere to reradiate more energy than it receives. It is thus impossible for any stellar-heating driven scenario to explain the observed over-luminosity given that CWW 89Ab intercepts a total of $\log(\mathrm{L/\mathrm{L}_\odot})=-4.85$ from the star CWW 89A. 
\\ \newline
3) \emph{Tidal-heating driven luminosity:} The combination of the short period and the non-zero eccentricity of CWW 89Ab's orbit indicate that the brown dwarf is still in the process of tidally circularizing about CWW 89A. As a result, there must be some additional energy being deposited within the interior of CWW 89Ab by tidal heating. Predicting the exact amount of tidal heating in both brown dwarfs and hot Jupiters is a task fraught with uncertainties (though see Section 5), so for now let us simply consider the overall energy budget of CWW 89Ab's orbit.

Effectively all the tidal heating energy will come from the orbital energy of CWW 89Ab. Recall that for a two-body system the zero-point for orbital energy is a perfectly parabolic, unbound, orbit with an eccentricity of unity, and that all bound orbits have negative energy values. For CWW 89Ab, the maximum amount of energy available for tidal heating will thus be the difference between its current orbital energy and a parabolic orbit with energy zero, or
\begin{equation}\label{eq:4110}
E_{orb}=\frac{\mathrm{G}\, M_A\, M_{BD}}{2a}=5.2\times10^{38}\,\mathrm{J},
\end{equation}
where G is the gravitational constant and $a$ is the current orbital semimajor axis. The average tidal heating rate is then simply $E_{orb}$ divided by the tidal time. For a system age of 3.0 Gyr, this gives an average tidal heating rate of $5.50\times10^{21}$\,W, or $\log(\mathrm{L_{tide}/\mathrm{L}_\odot})=-4.84$, which is again too low to explain the observed over-luminosity of $\log(\mathrm{L_{bol}/\mathrm{L}_\odot})=-4.19$.

On the other hand, if the tidal time is instead 0.65 Gyr, then the average heating rate would match the observed over-luminosity in CWW 89Ab. This short tidal evolution might occur if, for example, CWW 89Ab formed independently in the cluster and was recently captured by the CWW 89 binary. Though this sort of capture scenario is extremely unlikely \citep{bodenheimer2011}, the real problem with this short a tidal evolution time -- and generically with any sort of energy deposition deep within the interior of CWW 89Ab -- is that it would cause the radius of CWW 89Ab to be much larger than observed. As shown by the orange point from the Burrows models in Figure \ref{massradius}, an additional internal luminosity of $\log(\mathrm{L_{add}/\mathrm{L}_\odot})=-4$ would cause the radius of CWW 89Ab to inflate to approximately 1.3\rj, which is nearly 10\,$\sigma$ higher than observed. The timescale for the radius to inflate like this would only be about 10 Myr \citep[][and E. Lopez, private communication]{lopez2016}. This would imply that we are seeing CWW 89Ab only a few Myr into its tidal evolution -- but it would be impossible for the brown dwarf to have been captured into anything close to its present short period nearly-circular orbit. Thus, an extremely short tidal evolution time, and resulting high tidal heating, is not a possible explanation for the observed over-luminosity. 

\subsection{Possible Explanations}

\begin{figure}[t]
\vskip 0.00in
\includegraphics[width=1.1\linewidth,clip]{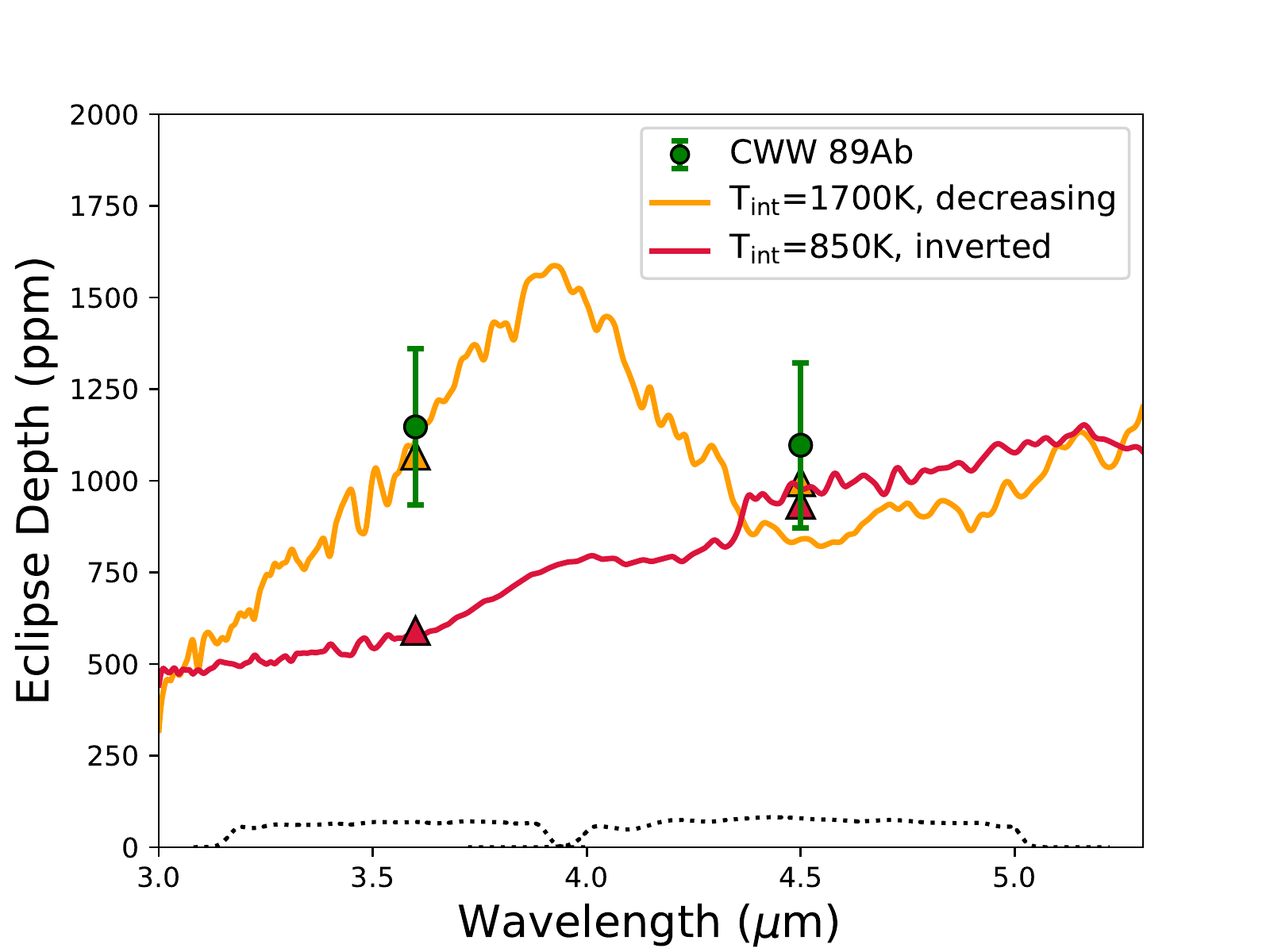}
\vskip -0.0in
\caption{One possible explanation for the observed over-luminosity on CWW 89Ab's dayside is the presence of a stratospheric temperature inversion. This would cause the deep \methane and \carbonmonoxide absorption features present in an atmosphere with a monotonically decreasing temperature to either vanish or become emission features. The inverted atmosphere model shown here (red line) is marginally consistent with the observed secondary eclipse depths, at $2.6\,\sigma$.}
\label{invert}
\vskip -0.05in
\end{figure}

1) \emph{A temperature inversion on CWW 89Ab's dayside:} All of the atmosphere models presented in Section 4 and Figure \ref{morley} assume that the temperature of CWW 89Ab's atmosphere monotonically decreases as pressure (and altitude) decreases. This causes the \methane and \carbonmonoxide bands at \three and \four to appear as absorption features in our models. If, on the other hand, the temperature-pressure profile of CWW 89Ab's dayside were inverted and had a region where temperature increased with decreasing pressure, then the \methane and \carbonmonoxide bands could appear as emission features (Figure \ref{invert}). 

In general, an atmospheric temperature inversion on the dayside of a hot Jupiter or irradiated brown dwarf requires the atmosphere to have strong absorption in the optical (near the peak of the stellar SED), and inefficient thermal cooling. The usual culprits behind temperature inversions in hot Jupiters are gas-phase TiO or VO in the upper atmosphere, which fulfill both criteria \citep{hubeny2003, fortney2008}. However, the dayside of CWW 89Ab of is cooler than the 1800K condensation temperature for both of these molecules, which therefore cannot be the cause of a temperature inversion.

Recently, however, \cite{molliere2015} suggested that an irradiated atmosphere with high surface gravity could display an inverted temperature-pressure profile if the C/O ratio in the atmosphere was near unity. For surface gravities at and above $\log(g)\sim5$ (cgs) pressure broadening causes the optical alkali lines to become very effective absorbers of the incoming stellar irradiation, which fulfills the first requirement for a temperature inversion. Making the atmospheric C/O$\approx$1 fulfills the second requirement, inefficient thermal cooling, by making the \carbonmonoxide dominate the carbon and oxygen budget of the atmosphere, inhibiting the formation of $\mathrm{CH}_4$, \water and HCN, and hence making it difficult for the atmosphere to reradiate energy away.

Using the same atmospheric models as in Section 3.5, we approximated the effect of a temperature inversion on CWW89Ab's dayside spectrum by adding an arbitrary $10^8$ erg cm$^{-2}$ s$^{-1}$ to the atmosphere at 0.1 bar (red line, Figure \ref{invert}). This caused the CO feature around 4.5\,\micron\ to invert into a mild emission feature, closely matching the observed \four secondary eclipse depth. At \threealt, the high temperatures caused by the inversion dissociated the atmospheric $\mathrm{CH}_4$, since our models assume chemical equilibrium. This caused the inverted \three secondary eclipse depth to be significantly lower than the observed value. Nevertheless, the inverted atmospheric model shown in Figure \ref{invert} is consistent with the observations, at $2.6\,\sigma$.
\\ \newline
2) \emph{Errors in brown dwarf evolutionary models:} Finally, there may simply be errors in the brown dwarf evolutionary models. As mentioned, CWW 89Ab is the first brown dwarf for which we have a directly measured mass, radius, age and luminosity. This makes it the first direct test of the physical parameters underlying the evolutionary models, and it would not be surprising for there to be some disagreement between observations and theory.

As noted in the Introduction, there have been indications that current brown dwarf models do not accurately describe the radius and luminosity evolution of these objects. Some of the other transiting brown dwarfs, such as KELT-1b, have radii that are significantly inflated above model predictions \citep{beatty2017}. The non-transiting brown dwarf WD0137-349B appears three times over-luminous in $K$ and at \four, though this may the result of dayside $\mathrm{H}_2$ florescence or $\mathrm{H}_3^+$ emission \citep{casewell2015}. Finally, dynamical mass measurements of the brown dwarf-brown dwarf binaries HD 130948BC \citep{dupuy2009} and Gl 417BC \citep{dupuy2014} show that all four of these objects are over-luminous by a factor of 1.6 relative to evolutionary models.

That being said, the over-luminosity in CWW 89Ab is 16 times the model predictions, which seems difficult to explain by model inaccuracies alone. This is because the radius of CWW 89Ab does appear to perfectly match model predictions (Figure \ref{massradius}). Therefore, to explain the observed over-luminosity, models of the effective temperature evolution of brown dwarfs would need to increase by a factor of two. Given the general agreement between evolutionary models and the ensemble color-magnitude diagrams of field brown dwarfs \citep[e.g.][]{sm08}, a factor of two increase seems hard to account for. Still, given that CWW 89Ab is the first time we have independent mass, radius, age, and luminosity measurements, an error in the evolutionary models is possible -- though we consider it unlikely.

\section{Constraints on the Tidal Quality Factor and Tidal Heating of CWW 89Ab}

In addition to measuring the dayside emission from CWW 89Ab, we are also able to place an interesting lower limit on the tidal quality factor of CWW 89Ab. This is due to a combination of two factors. First, our secondary eclipse measurements provide more precise values for the orbital eccentricity and argument of periastron than the radial velocity orbits measured in either \cite{nowak2017} or \cite{curtis2018}. Second, we have a strong expectation for what the rotation period of CWW 89A would be if it were in isolation, from gyrochronology (20.5 days, J. van Saders, private communication). As has been mentioned, CWW 89A has an observed photometric period of 12.66 days. The difference between this observed period and the expected gyrochronological period enables us to compute the amount of angular momentum transferred between CWW 89Ab's orbit and CWW 89A's rotation. Coupled, these two factors allow us to consider the tidal history of the CWW 89A system in more detail than was possible in \cite{nowak2017}.

To do so, we integrated the tidal evolution equations from \cite{leconte2010} to backwards in time from the current system age of 3.0 Gyr to 0.1 Gyr (Figure 12). We fixed the current orbital properties to the values listed in Table 4, and the stellar and brown dwarf masses to be those in Tables \ref{tab:stellarprops} and \ref{tab:priors}, respectively. Since we also wished to keep track of the angular momentum evolution of the star CWW 89A, we set the stellar moment of inertia to match that of the Sun: $I/M_*R_*^2=0.070$. We chose to ignore the rotational angular momentum of CWW 89Ab itself, since this will be negligible compared to the stellar rotational and CWW 89Ab's orbital angular momenta.

\begin{figure}[t]
\vskip 0.00in
\includegraphics[width=1.1\linewidth,clip]{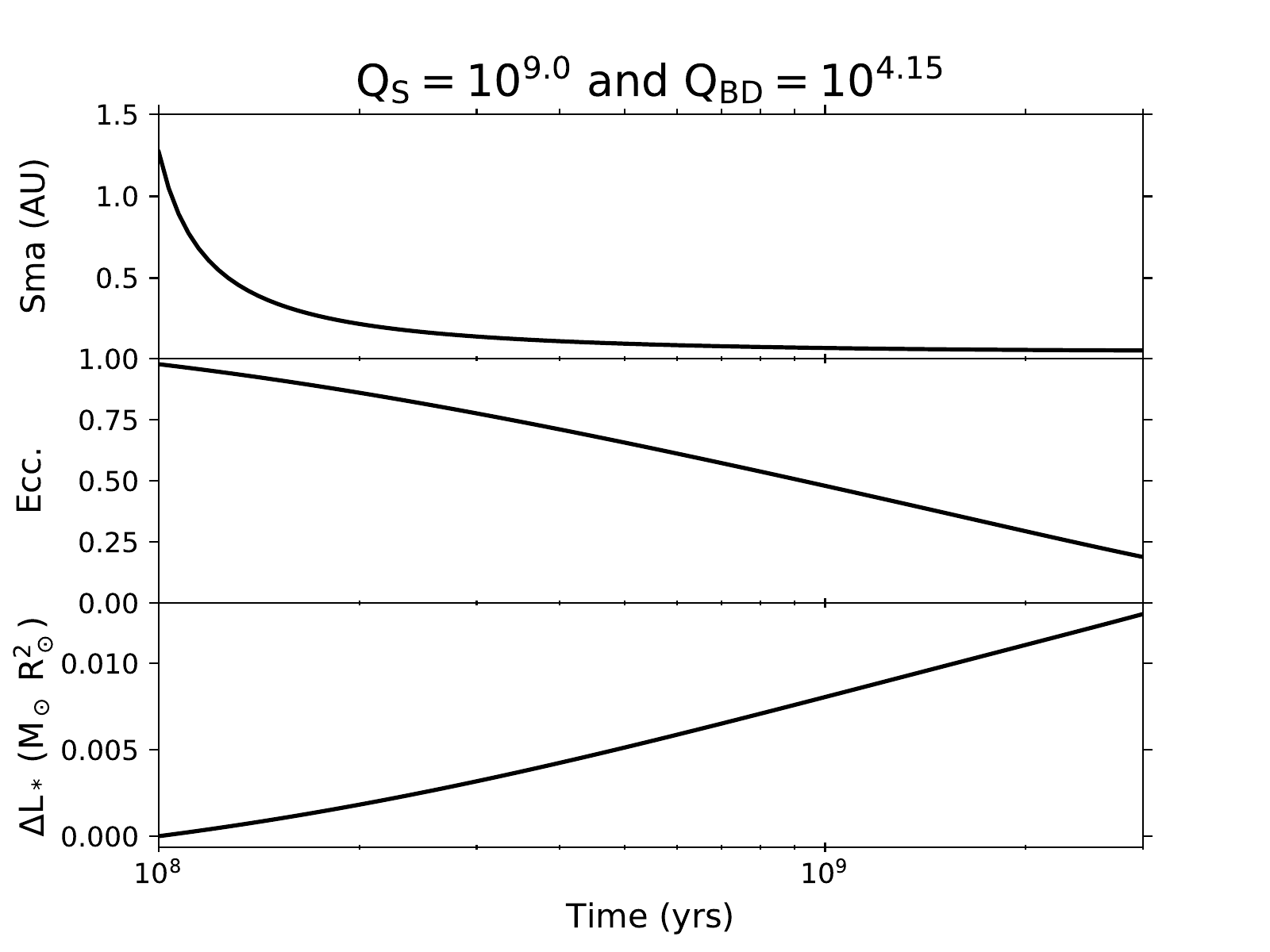}
\vskip -0.0in
\caption{The backwards time evolution of the orbit of CWW 89Ab and the rotation period of CWW 89A calculated using \cite{leconte2010}. If we require that the orbit of CWW 89Ab has $e<1$ and that CWW 89A gains rotational angular momentum of $\Delta L_* = 0.0132\,M_\odot\,R_\odot^2$, then we find that $Q_*>10^{9}$ and $Q_{BD}>10^{4.15}$. Here we show the limiting case that just fulfills the two physical conditions.}
\label{tides}
\vskip -0.05in
\end{figure}

This left us with two free parameters in our tidal evolution integration: the tidal quality factors for the star and the brown dwarf, $Q_*$ and $Q_{BD}$ \citep{Goldreich1963}. To determine the limits on both, we varied each value of $Q$ over a range of possible values from $10^3$ to $10^10$. For each trial tidal integration, we judged the run valid if the eccentricity of CWW 89Ab's orbit was less than one at the stopping point of 0.1 Gyr, and if the rotational angular momentum gained by CWW 89A was within 10\% of $\Delta L_* = 0.0132\,M_\odot\,R_\odot^2$. This latter quantity is the difference in angular momentum between CWW 89A's observed 12.66 day rotation period, and the gyrochronology prediction of 20.5 days. As expected, we found that the choice of $Q_*$ primarily influenced the evolution of the stellar angular momentum, and the choice of $Q_{BD}$ the brown dwarf's orbital evolution.

We find that $Q_*>10^{9}$ and $Q_{BD}>10^{4.15}$ are required to fulfill the physical conditions of a bound orbit for CWW 89Ab and the necessary amount of angular momentum transder for CWW 89A at 0.1 Gyr. This lower limit on the the value of $Q_{BD}$ is only the second time a constraint on the tidal quality factor of a brown dwarf has been made. The only other estimate of tidal $Q$ for a brown dwarf was made by \cite{heller2010} using the transiting brown dwarf-brown dwarf system 2MASS J05352184-0546085, where they found that tidal synchronization required $Q_{BD}>10^{4.5}$. Both of these lower limits are consistent with estimates for the tidal quality factors of M-dwarfs ($\sim10^5$) and that of Jupiter ($\sim10^6$), and thus cannot provide us with much information about the interior structure of CWW 89Ab.

Our lower limit on $Q_{BD}$ does allow us to perform a more detailed calculation of the possible tidal heating rate in CWW 89Ab than presented in Section 4.1. The maximum tidal heating will occur at the lowest possible value of $Q_{BD}$, so our determination of $Q_{BD}>10^{4.5}$ allows us to place a rough upper limit on the amount of tidal heating possible. To do so, we used the tidal heating equations from \cite{leconte2010} to compute the heating rate at each step of our tidal evolution integration. We find that the heating rate peaks at early-times around $1\times10^{22}$\,W, or $10^{-4.6}\,\mathrm{L}_\odot$. The present-day heating rate is considerably lower, at approximately $10^{-6.6}\,\mathrm{L}_\odot$. While the early-time heating is higher than the order-of-magnitude tidal heating rate we estimated in Section 5.1, the present-day heating rate is significantly lower than the observed over-luminosity in CWW 89Ab. We thus consider it unlikely that tidal heating is responsible for CWW 89Ab's over-luminosity. 

We caution the reader that our limits on $Q_{BD}>10^{4.5}$ and the tidal heating rate should not be considered accurate to more than a factor of a few. In these computations we have used a constant-phase-lag equilibrium tide model for the orbital evolution and heating of CWW 89Ab. In detail this model will be incorrect when the eccentricity of CWW 89Ab's orbit begins to approach unity, since on a highly eccentric orbit the tidal deformation of CWW 89Ab will become an ``impulsive'' event occurring only near periastron. That said, these non-equilibrium tides will only effect the evolution and heating of CWW 89Ab at early times, and we consider this effect negligible relative to the above uncertainties. Another aspect that we neglect is the possible time evolution of the stellar tidal quality factor at very early time \citep[1 to 10s of Myr,][]{heller2018}. Though we halt our integration at 100 Myr, it is possible that the time evolution of CWW 89A's $Q_*$ in the first few Myr may effect the ``initial'' oribtal conditions we require at 100 Myr.

Finally, the presence of CWW 89B raises the strong possibility that the orbit of CWW 89Ab has experienced Lidov-Kozai \citep{lidov1962,kozai1962} cycles at some point in the past, or even up to the present day. If CWW 89Ab was currently undergoing Lidov-Kozai oscillations the proceeding tidal evolution estimates would be invalid, and -- more importantly -- the rapidly cycling orbital eccentricity could provide a method to temporarily tidally heat CWW 89Ab without inflating its radius. However, given the relatively high mass of CWW 89Ab, General Relativistic precession will quench Lidov-Kozai oscillations once $a<0.4$\,AU \citep{fabrycky2007}. Since we expect the orbit of CWW 89Ab to shrink below this semi-major axis in a few hundred Myrs (R. Dawson, private communication), Lidov-Kozai oscillations should only occur very early in the system's history, and are impossible at the present day.

\section{Discussion and Summary}

We observed two secondary eclipses of the brown dwarf CWW 89Ab using Spitzer/IRAC at \three and \fouralt. After correcting for the dilution of the unresolved binary companion CWW 89B, we measure secondary eclipse depths of $\delta_{3.6}=1147\pm213$\,ppm and $\delta_{4.5}=1097\pm225$\,ppm. Accounting for the stellar irradiation that the dayside of CWW 89Ab receives from its host star, our atmospheric modeling implies that the internal heat in CWW 89Ab is therefore $\mathrm{T}_\mathrm{int}=1700\pm130$\,K. The internal luminosity of CWW 89Ab is thus $\log(\mathrm{L_{bol}/\mathrm{L}_\odot})=-4.19\pm0.14$, which is 16 times, or $9.3\,\sigma$, higher than predictions from irradiated or non-irradiated brown dwarf evolutionary models (Figure \ref{agelum}).

Since CWW 89Ab is the first transiting brown dwarf for which we have an independently measured mass, radius, age, and luminosity, the  over-luminosity is a strong challenge for these evolutionary models.

As we discuss in Section 4.1, the observed over-luminosity cannot be explained by an inaccurate age determination, stellar heating, or by tidal heating. Both the age and tidal explanations suffer from the problem that while they can, in specific scenarios, explain the luminosity of CWW 89Ab, both would necessitate that the radius of CWW 89Ab be $4.5\,\sigma$ larger than observed (Figure \ref{massradius}). Stellar heating explanations are unworkable because there is simply not enough energy incoming from the star.

In principle, it is possible to explain CWW 89Ab's over-luminosity by invoking a large error in brown dwarf evolutionary models. Since the radius of CWW 89Ab matches model predictions, this would require the predicted effective temperatures for objects near 36\mj\ and 3.0 Gyr to increase by a factor of two. At the same time, any model changes necessary to explain CWW 89Ab would also need to preserve the existing  agreement between the predicted and observed colors of field brown dwarfs. At first glance, this seems extremely challenging. We therefore consider an error in the evolutionary models to be very unlikely, but we cannot categorically rule it out.

Instead, we suggest that the most likely explanation is that the dayside atmosphere of CWW 89Ab has a stratospheric temperature inversion (Figure \ref{invert}). The atmosphere of CWW 89Ab is too cool for gas-phase TiO or VO to exist in quantity, so unlike in hot Jupiters a temperature inversion on CWW 89Ab's dayside would be caused by absorption from pressure broadened optical alkali lines. As described in \cite{molliere2015} this could occur if the atmospheric C/O$\approx$1. 

\subsection{Possible Implications for the Formation Pathway of CWW 89Ab}

If the atmospheric carbon-to-oxygen ratio of CWW 89Ab is indeed close to unity, this would have a strong implication for how CWW 89Ab formed. The elemental abundances of Ruprecht 147 stars are close to Solar \cite{curtis2013}, which implies that the star CWW 89A -- and the material from which it formed -- has a Solar-like C/O=0.5. If CWW 89Ab formed via direct gravitational collapse alongside CWW 89A, then it should also have C/O=0.5 \citep[e.g.,][]{madhu2017}.

A super-stellar C/O ratio of unity, on the other hand, would mean that the material used to form CWW 89Ab was first processed through CWW 89A's proto-planetary disk \citep{oberg2011}. In this scenario, it follows that CWW 89Ab would then have formed via core accretion within the proto-planetary disk, and not through gravitational collapse. At 36.5\mj\ this would make CWW 89Ab the most massive object known to from via core accretion, and it would represent the very tail end of the expected mass distribution from this formation pathway \citep{ida2004, mordasini2012}.

The positive identification of CWW 89Ab's formation pathway as core accretion, rather than gravitational collapse, would also make CWW 89Ab one of two brown dwarfs with a defined formation pathway. \cite{dupuy2018} recently suggested that 2MASS J0249-0557c is an 11.6\mj\ object that formed via gravitational collapse, which would make the confirmation that CWW 89Ab formed through core accretion an interesting comparison. This would provide a unique high-mass anchor for core accretion theories of planet formation, and could provide some insight into the reasons behind the the brown dwarf desert \citep{grether2006}.
\newline

We reiterate though, that the possibility that CWW 89Ab has an atmospheric C/O$\approx$1 is contingent on our inference that a temperature inversion exists on CWW 89Ab's dayside. This inference, that a dayside inversion is present, could be directly tested in two ways. First, a JWST eclipse spectrum with NIRCam or NIRSpec covering the IRAC wavelengths would directly show wjhether or not the \methane and \carbonmonoxide features are truly in emission, and inverted. Second, since an inversion can only occur on the dayside atmosphere, broadband Spitzer/IRAC observations of the nightside emission -- if they were consistent with the predicted internal luminosity of CWW 89Ab -- would also demonstrate that the dayside atmosphere is inverted.

\acknowledgements

The authors would like to thank Jennifer van Saders and Rebekah Dawson for helpful discussions during the drafting process.

This work is based on observations made with the Spitzer Space Telescope, which is operated by the Jet Propulsion Laboratory, California Institute of Technology under a contract with NASA. Support for this work was provided by NASA. This work was partially funded by the National Aeronautics and Space Administration under grant NNX16AE64G issued through the \textit{K2} Guest Observer Program (GO 7035).

T. G. Beatty was partially supported by funding from the Center for Exoplanets and Habitable Worlds. The Center for Exoplanets and Habitable Worlds is supported by the Pennsylvania State University, the Eberly College of Science, and the Pennsylvania Space Grant Consortium. J. L. Curtis was supported by the National Science Foundation Astronomy and Astrophysics Postdoctoral Fellowship under award AST-1602662. Work by B. T. Montet was performed under contract with the Jet Propulsion Laboratory (JPL) funded by NASA through the Sagan Fellowship Program executed by the NASA Exoplanet Science Institute. J.R.A. Davenport is supported by an NSF Astronomy and Astrophysics Postdoctoral Fellowship under award AST-1501418.

This work has made use of NASA's Astrophysics Data System, the Exoplanet Orbit Database and the Exoplanet Data Explorer at exoplanets.org \citep{exoplanetsorg}, the Extrasolar Planet Encyclopedia at exoplanet.eu \citep{exoplanetseu}, the SIMBAD database operated at CDS, Strasbourg, France \citep{simbad}, and the VizieR catalog access tool, CDS, Strasbourg, France \citep{vizier}. 

\facility{Spitzer (IRAC)}

\software{Astropy \citep{astropy}, BATMAN \citep{kreidberg2015}, emcee \citep{dfm2013}}

\end{document}